\documentclass[twocolumn]{aastex63} 

\usepackage[caption = false]{subfig}
\usepackage{graphicx}

\begin{document}

\title{Imaging the water snowline around protostars with water and HCO$^+$ isotopologues}

\correspondingauthor{Merel L.R. van 't Hoff}
\email{mervth@umich.edu}

\author{Merel L.R. van 't Hoff}
\affil{Leiden Observatory, Leiden University, P.O. Box 9513, 2300 RA Leiden, The Netherlands}
\affil{Department of Astronomy, University of Michigan, 1085 S. University Ave., Ann Arbor, MI 48109-1107, USA}

\author{Daniel Harsono} 
\affil{Leiden Observatory, Leiden University, P.O. Box 9513, 2300 RA Leiden, The Netherlands}
\affil{Institute of Astronomy and Astrophysics, Academia Sinica, P.O. Box 23-141, Taipei 106, Taiwan} 

\author{Martijn L. van Gelder} 
\affil{Leiden Observatory, Leiden University, P.O. Box 9513, 2300 RA Leiden, The Netherlands}

\author{Tien-Hao Hsieh}
\affil{Institute of Astronomy and Astrophysics, Academia Sinica, P.O. Box 23-141, Taipei 106, Taiwan} 
\affil{Max-Planck-Institut f\"ur Extraterrestrische Physik, Giessenbachstrasse 1, 85748 Garching, Germany}

\author{John J. Tobin}
\affil{National Radio Astronomy Observatory, 520 Edgemont Rd., Charlottesville, VA 22903, USA}

\author{Sigurd S. Jensen}
\affil{Niels Bohr Institute, University of Copenhagen, {\O}ster Voldgade 5-7, 1350 Copenhagen K., Denmark}
\affil{Max-Planck-Institut f\"ur Extraterrestrische Physik, Giessenbachstrasse 1, 85748 Garching, Germany}

\author{Naomi Hirano}
\affil{Institute of Astronomy and Astrophysics, Academia Sinica, P.O. Box 23-141, Taipei 106, Taiwan}

\author{Jes K. J{\o}rgensen}
\affil{Niels Bohr Institute, University of Copenhagen, {\O}ster Voldgade 5-7, 1350 Copenhagen K., Denmark}

\author{Edwin A. Bergin}
\affil{Department of Astronomy, University of Michigan, 1085 S. University Ave., Ann Arbor, MI 48109-1107, USA}

\author{Ewine F. van Dishoeck}
\affil{Leiden Observatory, Leiden University, P.O. Box 9513, 2300 RA Leiden, The Netherlands}
\affil{Max-Planck-Institut f\"ur Extraterrestrische Physik, Giessenbachstrasse 1, 85748 Garching, Germany}




\begin{abstract}

\noindent The water snowline location in protostellar envelopes provides crucial information about the thermal structure and the mass accretion process as it can inform about the occurrence of recent ($\lesssim$1,000 yr) accretion bursts. In addition, the ability to image water emission makes these sources excellent laboratories to test indirect snowline tracers such as H$^{13}$CO$^+$. We study the water snowline in five protostellar envelopes in Perseus using a suite of molecular line observations taken with the Atacama Large Millimeter/submillimeter Array (ALMA) at $\sim$0.2$^{\prime\prime}-$0.7$^{\prime\prime}$ (60--210 au) resolution. B1-c provides a textbook example of compact H$_2^{18}$O ($3_{1,3}-2_{2,0}$) and HDO ($3_{1,2}-2_{2,1}$) emission surrounded by a ring of H$^{13}$CO$^+$ ($J=2-1$) and HC$^{18}$O$^+$ ($J=3-2$). Compact HDO surrounded by H$^{13}$CO$^+$ is also detected toward B1-bS. The optically thick main isotopologue HCO$^+$ is not suited to trace the snowline and HC$^{18}$O$^+$ is a better tracer than H$^{13}$CO$^+$ due to a lower contribution from the outer envelope. However, since a detailed analysis is needed to derive a snowline location from H$^{13}$CO$^+$ or HC$^{18}$O$^+$ emission, their true value as snowline tracer will lie in the application in sources where water cannot be readily detected. For protostellar envelopes, the most straightforward way to locate the water snowline is through observations of H$_2^{18}$O or HDO. Including all sub-arcsecond resolution water observations from the literature, we derive an average burst interval of $\sim$10,000 yr, but high-resolution water observations of a larger number of protostars is required to better constrain the burst frequency.  

\end{abstract}




\section{Introduction} \label{sec:intro}

Young stars are surrounded by disks of dust, gas, and ice. The location in the disk where the transition between gas and ice occurs is molecule dependent, and is set by the species-specific binding energy to the grains and the temperature structure in the circumstellar material. The sequential freeze out of molecules at their so-called snowlines creates a chemical gradient in the gas and ice, and the composition of forming planets is thus related to the location where they accrete most of their solids and gas \citep[e.g.,][]{Oberg2011,Madhusudhan2014,Walsh2015,Mordasini2016,Eistrup2016,Booth2017,Cridland2019}.  In addition, the growth of dust grains, and thus the planet formation efficiency, is thought to be significantly enhanced at the water snowline \citep[e.g.,][]{Stevenson1988,Schoonenberg2017,Drazkowska2017}. 

Unfortunately, it is very challenging to observe the water snowline in protoplanetary disks. Because of the large binding energy of water, freeze-out already occurs at temperatures $\sim$100--150 K, that is, at radii of a few au in disks around Sun-like stars ($\sim$0.01$^{\prime\prime}$ for nearby star-forming regions). Moreover, only emission from the less abundant isotopologue H$_2^{18}$O can be observed from the ground (except for H$_2$O lines that are often masing and generally tracing shocks). Another complication is that gas-phase water may not be as abundant in disks as expected from interstellar ice abundances \citep[e.g.,][]{Hogerheijde2011,Zhang2013,Du2017}. Observations of warm water in disks therefore require both high angular resolution and high sensitivity, and as such, only one spatially unresolved detection has been made so far \citep{Carr2018,Notsu2019}.

Younger disks that are still embedded in their envelope are warmer than mature protoplanetary disks \citep{vantHoff2018b,vantHoff2020}, and are expected to have their water snowline at larger radii \citep{Harsono2015}. However, no water emission was detected in a sample of five Class I disks and upper limits for the water abundance are one to three orders of magnitude lower than the interstellar ice abundance \citep{Harsono2020,vanDishoeck2021}. In addition, the non-detections of methanol, which desorbes at a similar temperature as water, toward a sample of Class I disks in both Taurus and Ophiuchus suggest that these sources do not have large hot ($\gtrsim$100--150 K) inner regions \citep{ArturdelaVillarmois2019,vantHoff2020}. 

So far, resolved water observations only exist for protostellar envelopes (possibly with a central disk-like structure) around Class 0 sources \citep{Jorgensen2010,Persson2012,Persson2013,Taquet2013,Bjerkeli2016,Jensen2019}. These objects have therefore been used to test the application of HCO$^+$ as chemical tracer of the water snowline, which may then be applied in sources such as disks where water is not readily observable. HCO$^+$ is expected to be a good snowline tracer, because its most abundant destroyer in the warm dense gas around young stars is gaseous H$_2$O. HCO$^+$ is therefore expected to be abundant only in the region where water is frozen out and gaseous CO is available for its formation \citep[][]{Phillips1992,Bergin1998}. 

The first observational hint that HCO$^+$ can trace the water snowline came from observations of the Class 0 protostar IRAS 15398--3359. The optically thin isotopologue H$^{13}$CO$^+$ displays ring-shaped emission in this source while the methanol emission is centrally peaked \citep{Jorgensen2013}. The distribution of HDO emission is complicated due to its presence along the outflow cavity wall but the central component lies within the H$^{13}$CO$^+$ ring \citep{Bjerkeli2016}. Subsequent observations of the spatial anticorrelation between H$_2^{18}$O and H$^{13}$CO$^+$ in the Class~0 protostar NGC1333-IRAS2A provided a proof of concept that H$^{13}$CO$^+$ can be used to trace the water snowline \citep{vantHoff2018a}. Recently, the value of H$^{13}$CO$^+$ was demonstrated by constraining the snowline location in the disk around the outbursting young star V883 Ori \citep{Leemker2021}.  

Locating the water snowline in protostellar envelopes is interesting by itself, because such observations can be used to trace episodic accretion \citep{Audard2014}; a snowline location at larger radii than expected from the source luminosity indicates that the protostar may have recently undergone an accretion burst \citep{Lee2007,Visser2012}. During the burst, the luminosity increases, heating up the circumstellar material and shifting the snowlines outward. While the temperature adapts almost instantaneously when the protostar goes back to its quiescent mode of accretion \citep{Johnstone2013}, the chemistry needs time to react. Hence molecules can remain in the gas phase out to larger radii than expected from the current luminosity. 

This concept was applied to C$^{18}$O observations, which suggested that protostars undergo a significant burst every 20,000--50,000 year \citep{Jorgensen2015,Frimann2017}. Since the water snowline will shift back faster after a burst than the CO snowline due to higher densities (100--1,000 yr versus $\sim$10,000 yr; \citealt{Visser2012}), observations of the water snowline could place more stringent constraints on the burst frequency and are therefore crucial for gaining a better understanding of the mass accretion process. Based on observations of N$_2$H$^+$ and HCO$^+$ as tracers of the CO and water snowline, respectively, \citet{Hsieh2019} derived burst intervals of 2400 yr during the Class 0 stage and 8000 yr during the Class I stage. 

In this work, we study the water snowline in five protostellar envelopes (B1-bS, B1-c, B5-IRS1, HH211 and L1448-mm) using dedicated and archival observations with the Atacama Large Millimeter/submillimeter Array (ALMA) of H$_2^{18}$O, HDO, HCO$^+$, H$^{13}$CO$^+$, and HC$^{18}$O$^+$. The goals are threefold: 1) directly locate the water snowline with H$_2^{18}$O (B1-c, HH211) and/or HDO emission (B1-bS, B1-c); 2) determine which HCO$^+$ isotopologues and transitions are best suited to trace the water snowline (B1-c); 3) determine whether these protostars have recently undergone an accretion burst, as well as the average burst interval. 

The paper is organized as follows. First, the observations are described in Sect.~\ref{sec:Observations}. We then present and discuss the H$_2^{18}$O $3_{1,3}-2_{2,0}$, HDO $3_{1,2}-2_{2,1}$ (Sect.~\ref{sec:B1-c_water}), H$^{13}$CO$^+$ $J=2-1$ and $J=3-2$ (Sect.~\ref{sec:B1-c_H13CO+}), HCO$^+$ $J=3-2$ and HC$^{18}$O$^+$ $J=3-2$ (Sect.~\ref{sec:B1-c_isotopologues}) observations toward B1-c. Next, the HDO $3_{1,2}-2_{2,1}$ and H$^{13}$CO$^+$ $J=3-2$ observations toward B1-bS are shown in Sect.~\ref{sec:B1-bS}, followed by the H$^{13}$CO$^+$ $J=2-1$ observations toward B5-IRS1, HH211 and L1448-mm in Sects.~\ref{sec:HH211}-\ref{sec:B5-IRS1}. Accretion bursts are discussed in Sect.~\ref{sec:Bursts}, where we first discuss the recent occurrence of a burst in the sources in our sample (Sect.~\ref{sec:Bursts_H13CO+sources}) and then derive the frequency of accretion bursts using all available sub-arcsecond resolution water observations (Sect.~\ref{sec:Bursts_Watersources}). Finally, Sect.~\ref{sec:Conclusions} summarizes the main conclusions.


\section{Observations} \label{sec:Observations} 

We have targeted four of the more luminous protostars in the Perseus molecular cloud ($d \sim$300 pc; \citealt{Ortiz-Leon2018}) that do not have a close ($< 4^{\prime\prime}$) companion \citep{Tobin2016}: B1-c, B5-IRS1, HH211 and L1448-mm (see Table~\ref{tab:OverviewSources}). L1448-mm is in a 8.1$^{\prime\prime}$ binary \citep{Jorgensen2006}. The H$^{13}$CO$^+$ $J=2-1$ transition was observed with ALMA (project code 2017.1.01371.S) for a total on source time of 23 minutes per source. In addition to spectral windows with 61 kHz ($\sim$0.1 km s$^{-1}$) resolution, the correlator setup included two continuum windows with 977 kHz (1.6--1.7 km~s$^{-1}$) spectral resolution centered at 174.106 and 187.493 GHz. Observations of the H$_2^{18}$O $3_{1,3}-2_{2,0}$ and HDO $3_{2,1}-4_{0,4}$ transitions were only taken toward B1-c and HH211 (ALMA project code 2019.1.00171.S). The total on source time was 36 minutes per source. The correlator setup again contains spectral windows with 61 kHz ($\sim$0.09 km~s$^{-1}$) resolution and two continuum windows with 977 kHz ($\sim$1.4 km~s$^{-1}$) resolution centered at 204.200 and 206.200 GHz. For both datasets, calibration was done using the ALMA Pipeline and versions 5.1.1 and 5.6.1, respectively, of the Common Astronomy Software Applications (CASA; \citealt{McMullin2007}). Self-calibration on the continuum as well as imaging was done using CASA version 5.6.1. To obtain the best image quality, the data were imaged using natural weighting. The maximum resolvable scale of the H$^{13}$CO$^+$ and H$_2^{18}$O data is $\sim$5$^{\prime\prime}$ and $\sim$7$^{\prime\prime}$, respectively. We may thus not recover H$^{13}$CO$^+$ emission from the outermost envelope, but our focus here is the inner few hundred au.   

B1-c was also observed as part of ALMA project 2017.1.01693.S, covering HCO$^+$ $J=3-2$ and HC$^{18}$O$^+$ $J=3-2$ at 30.5 kHz and 61 kHz ($\sim$0.03 and $\sim$0.07 km~s$^{-1}$) resolution, respectively. In addition, the H$^{13}$CO$^+$ $J=3-2$ transition was covered by ALMA project 2017.1.01174.S at 122 kHz ($\sim$0.14 km~s$^{-1}$) resolution. The reduction of these datasets are described in \citet{Hsieh2019} and \citet{vanGelder2020}, respectively. Both programs also targeted the protostar B1-b and the protostar B1-bS is present near the edge of the primary beam of these observations. None of the lines discussed here were detected toward B1-b. 

Observations of the HDO $3_{1,2}-2_{2,1}$ transition toward B1-c, B1-bS and B1-bN are present in the ALMA archive (project 2016.1.00505.S), and imaged as part of the ARI-L project \citep{Massardi2019}\footnote{https://almascience.org/alma-data/aril} using the ALMA Pipeline and CASA 5.6.1. Upon inspection of the weblog and the data, we have decided that these images can be used directly. 

\begin{deluxetable}{llcccc}
\tablecaption{Overview of sources. \label{tab:OverviewSources}}
\tablewidth{0pt}
\addtolength{\tabcolsep}{-3pt} 
\tabletypesize{\scriptsize}
\tablehead{
\colhead{Source\tablenotemark{a}} \vspace{-0.3cm} & \colhead{Other name}  & \colhead{R.A.\tablenotemark{b}} & \colhead{Dec.\tablenotemark{b}} & \colhead{Class} & \colhead{$L_{\rm{bol}}$\tablenotemark{c}}  \\ 
\colhead{} \vspace{-0.5cm}& \colhead{} & \colhead{(J2000)} & \colhead{(J2000)} & \colhead{} & \colhead{($L_{\odot}$)}  \\ 
} 
\startdata 
Per-emb-1 			& HH211-mms 	& 03:43:56.8 	& +32:00:50.2 		& 0 	& 3.6  \\
Per-emb-26 			&  L1448-mm 	& 03:25:38.9 	& +30:44:05.3 		& 0 	& 9.0 \\
Per-emb-29 			&  B1-c 				& 03:33:17.9 	& +31:09:31.8 		& 0 	& 5.2  \\
Per-emb-41 			&  B1-b 			& 03:33:20.3 	& +31:07:21.3 		& I 	& 0.3 \\
\nodata 				& B1-bN 			& 03:33:21.2		& +31:07:43.6		& 0 	& 0.3 \\
\nodata 				& B1-bS 			& 03:33:21.4 	& +31:07:26.3 		& 0 	& 0.5  \\
Per-emb-53 			&  B5-IRS1 		& 03:47:41.6 	& +32:51:43.7 		& I 	& 7.7  
\enddata
\vspace{0.1cm}
\textbf{Notes.}
Not all sources are observed in each molecular line, see Table~\ref{tab:OverviewLines} for an overview of molecular lines observed per source. 
\vspace{-0.2cm}
\tablenotetext{a}{Naming scheme of \citet{Enoch2009}.}
\vspace{-0.3cm} 
\tablenotetext{b}{Position of the continuum peak.}
\vspace{-0.3cm} 
\tablenotetext{c}{Luminosities for B1-bN and B1-bS are from \citet{Hirano2014}. For the other sources, when available, luminosities are taken from \citet{Karska2018} and otherwise they are taken from \citet{Tobin2016}. In all cases, luminosities are converted to a distance of 300 pc for Perseus. \citep{Ortiz-Leon2018}.}
\end{deluxetable}

\begin{deluxetable*}{lclccccc}
\tablecaption{Overview of molecular line observations. \label{tab:OverviewLines}}
\tablewidth{700pt}
\tabletypesize{\scriptsize}
\tablehead{
\colhead{Molecular line} \vspace{-0.3cm} & \colhead{Frequency} & \colhead{Source} & \colhead{Beam size} & \colhead{$\Delta v$} & \colhead{rms\tablenotemark{a}} &  \colhead{$F_{\rm{int}}$\tablenotemark{b}} & \colhead{ALMA} \\ 
\colhead{} \vspace{-0.5cm} & \colhead{(GHz)} & \colhead{} & \colhead{(arcsec)} & \colhead{(km s$^{-1}$)} & \colhead{(mJy beam$^{-1}$)} &  \colhead{(Jy km s$^{-1}$)} & \colhead{project code} \\ 
} 
\startdata 
HCO$^+$ $J=3-2$ & 267.557626  & B1-b & 0.46 $\times$ 0.30 & 0.10 & 5.2 &  $-$ & 2017.1.01693.S\\
                                  &                       & B1-c & 0.46 $\times$ 0.30 & 0.10 & 9.9 & 29.1 $\pm$ 0.6 &  \\
H$^{13}$CO$^+$ $J=2-1$ & 173.506700 &  B1-c & 0.72 $\times$ 0.60 & 0.11 & 6.7 & 5.94 $\pm$ 0.05 & 2017.1.01371.S\\
                                               &                      & B5-IRS1 & 0.77 $\times$ 0.60 & 0.11 & 6.7 & 2.92 $\pm$ 0.05 &  \\
                                               &                      & HH211 & 0.73 $\times$ 0.58 & 0.11 & 6.7 & 4.65 $\pm$ 0.05 &  \\
                                               &                      & L1448-mm & 0.70 $\times$ 0.59 & 0.11 & 6.7 & 5.91 $\pm$ 0.05 &  \\
H$^{13}$CO$^+$ $J=3-2$ & 260.255339 & B1-b & 0.58 $\times$ 0.39 & 0.14 & 2.7 & $-$ & 2017.1.01174.S\\
                                               &                       & B1-bS & 0.58 $\times$ 0.39 & 0.14 & 9.0 & 0.57 $\pm$ 0.02 &  \\
                                               &                       & B1-c & 0.58 $\times$ 0.39 & 0.14 & 2.5 & 12.52 $\pm$ 0.06 &  \\
HC$^{18}$O$^+$ $J=3-2$ & 255.479389 & B1-b & 0.48 $\times$ 0.33 & 0.10 & 6.0 & $-$ & 2017.1.01693.S\\
                                               &                       & B1-c & 0.48 $\times$ 0.33 & 0.10 & 6.0 & 1.86 $\pm$ 0.06 & \\
H$_2^{18}$O $3_{1,3}-2_{2,0}$ & 203.407520 & B1-c & 0.94 $\times$ 0.58 & 0.09 & 4.7 & 0.13 $\pm$ 0.01 & 2019.1.00171.S\\
                                                      &                       & HH211 & 1.09 $\times$ 0.68 & 0.09 & 4.0 & $-$ &  \\
HDO $3_{2,1}-4_{0,4}$              & 207.110852   & B1-c & 0.91 $\times$ 0.55 & 0.09 & 6.9 & 0.05 $\pm$ 0.01\tablenotemark{c}& 2019.1.00171.S \\   
                                                     &                       & HH211 & 1.06 $\times$ 0.65 & 0.09 & 6.0 &  $-$ &  \\      
HDO $3_{1,2}-2_{2,1} $ 			& 225.896720 & B1-bN & 0.25 $\times$ 0.16 & 0.16 & 2.2 & $-$ & 2016.1.00505.S  \\    
												&						& B1-bS & 0.25 $\times$ 0.15 & 0.16 & 2.2 & 0.03 $\pm$ 0.01 & \\
												&						& B1-c & 0.25 $\times$ 0.16 & 0.16 & 2.2 & 0.24 $\pm$ 0.01 &                                                                                            
\enddata
\tablenotetext{a}{Rms in channels with width $\Delta v$.}
\vspace{-0.3cm}
\tablenotetext{b}{Fluxes are extracted in a circular aperture with a diameter of 10$^{\prime\prime}$, except for B1-bS (3$^{\prime\prime}$), and the H$_2^{18}$O and HDO observations (derived using the CASA \textit{imfit} procedure). }
\vspace{-0.3cm}
\tablenotetext{c}{Most likely to be from an unidentified molecular line instead, see Sect.~\ref{sec:B1-c_water}.}
\end{deluxetable*}

Finally, to provide a full overview of the spatial extent of water emission observed to date at sub-arcsecond resolution toward protostars, we use the water observations (H$_2^{18}$O $3_{1,3}-2_{2,0}$, HDO $3_{1,2}-2_{2,1}$ and $2_{1,1}-2_{1,2}$, and D$_2$O $1_{1,0}-1_{0,1}$) toward the isolated protostars B335, L483 and BHR71-IRS1 (ALMA programs 2017.1.00693.S and 2019.1.00720.S). These data have been presented by \citet{Jensen2019} and \citet{Jensen2021}, but the source size was not reported in these works.

More observational details for all observing campaigns (including observing dates and calibrators) can be found in Table~\ref{tab:Observations}. Information on the observed molecular lines (including beam size and sensitivity) is listed in Table~\ref{tab:OverviewLines}. Continuum images for the protostellar envelopes (at 1.2 mm for B1-bS and at 1.7 mm for B1-c, B5-IRS1, HH211 and L1448-mm) are presented in Fig.~\ref{fig:Continuum}.


\section{Imaging the water snowline in the protostellar envelope of B1-c} \label{sec:B1-c}

\subsection{H$_2$\textsuperscript{18}O and HDO} \label{sec:B1-c_water}

\begin{figure*}
\centering
\includegraphics[width=\textwidth,trim={0.2cm 14.4cm 0cm 0.5cm},clip]{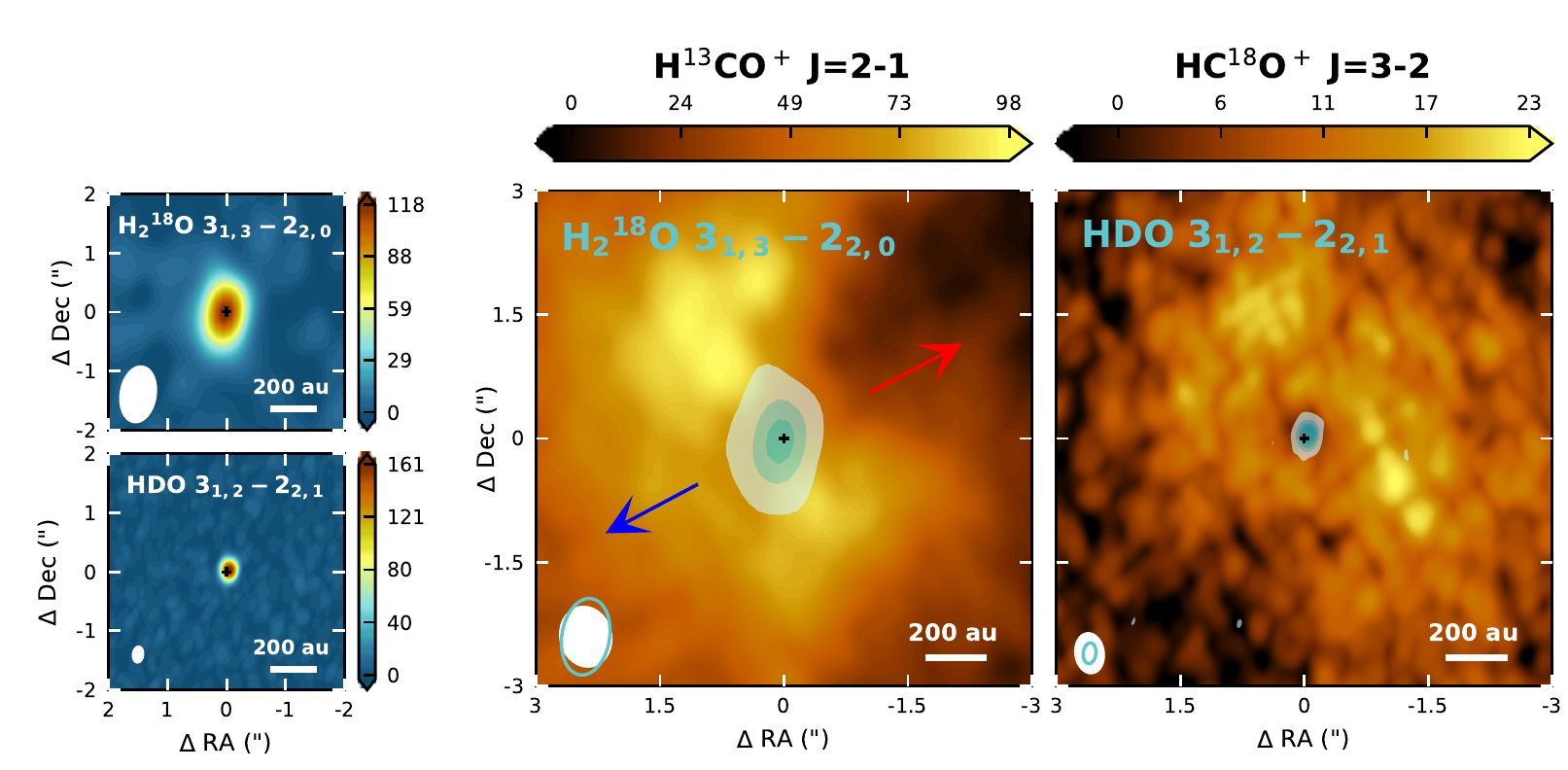}
\caption{Integrated intensity maps for the H$_2^{18}$O $3_{1,3}-2_{2,0}$ (top left) and HDO $3_{1,2}-2_{2,1}$ (bottom left) transitions toward B1-c. The H$_2^{18}$O and HDO integrated intensity maps are also overlaid in transparent shades of blue onto the H$^{13}$CO$^+$ $J=2-1$ (middle) and HC$^{18}$O$^+$ $J=3-2$ (right) integrated intensity maps, respectively. The displayed pairing is based on comparable beam sizes. In the overlaid images, the water images are clipped at $3\sigma$. For H$_2^{18}$O (HDO) the contours are at [3, 13, 23]$\sigma$ ([3, 13, 23, 33]$\sigma$), where $\sigma = 4.0$ (3.3) mJy beam $^{-1}$ km s$^{-1}$. All color scales are in mJy beam$^{-1}$ km s$^{-1}$. The continuum peak is marked by a cross and the beams are depicted in the lower left corners (blue contour for H$_2^{18}$O and HDO in the overlaid images). The outflow direction is indicated by blue and red arrows in the middle panel.}
\label{fig:M0_H13COp-H218O}
\end{figure*}

\begin{figure}
	\centering
	\subfloat{\includegraphics[trim={0.5cm 15.9cm 8cm 1.6cm},clip]{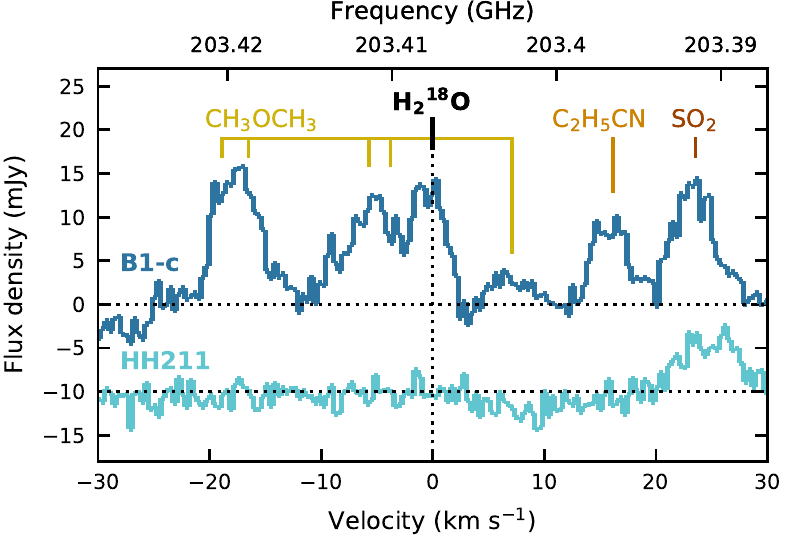}}
	\hfill
	\subfloat{\includegraphics[trim={0.5cm 15.9cm 8cm 1.6cm},clip]{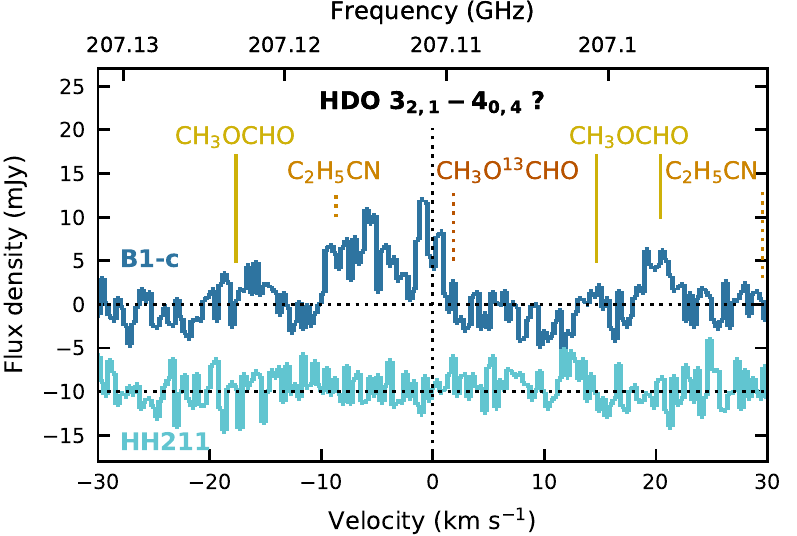}}
	\hfill
	\subfloat{\includegraphics[trim={0.5cm 15.9cm 8cm 1.6cm},clip]{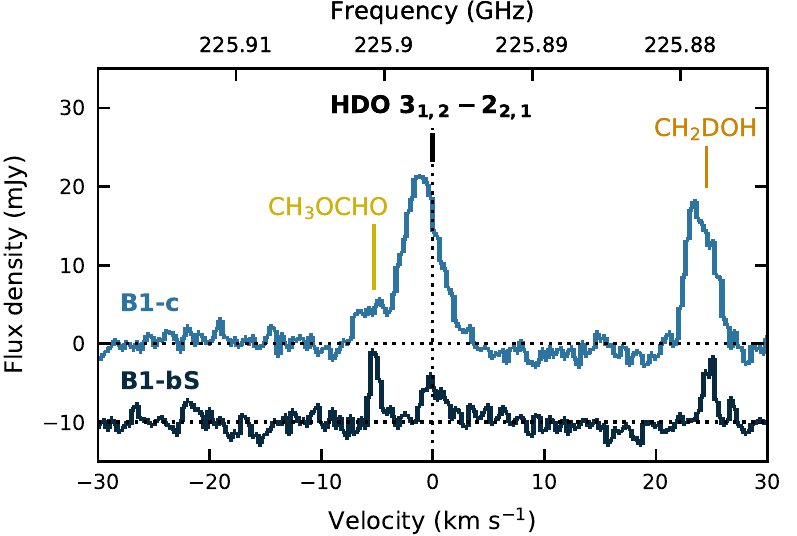}}
	\caption{Spectra toward B1-c (blue) and HH211 (light blue) centered on the H$_2^{18}$O $3_{1,3}-2_{2,0}$ (top) and HDO $3_{2,1}-4_{0,4}$ (middle) transitions, and toward B1-c (blue) and B1-bS (dark blue) centered on the HDO $3_{1,2}-2_{2,1}$ transition (bottom). The spectra are extracted within one beam toward the continuum position (see Table~\ref{tab:OverviewLines}) and binned to a resolution of 0.36 km s$^{-1}$ (H$_2^{18}$O and HDO $3_{2,1}-4_{0,4}$) or 0.32 km s$^{-1}$ (HDO $3_{1,2}-2_{2,1}$). The spectra for HH211 and B1-bS have a $-10$ mJy offset. Molecules labeled by a dotted line are not detected (the two C$_2$H$_5$CN lines in the middle panel are expected to have equal strenght). }
	\label{fig:SpectraH218O}
\end{figure}

Figure~\ref{fig:M0_H13COp-H218O} presents integrated intensity maps revealing compact, centrally peaked H$_2^{18}$O $3_{1,3}-2_{2,0}$ and HDO $3_{1,2}-2_{2,1}$ emission toward B1-c. Spectra extracted in the central beam are presented in Fig.~\ref{fig:SpectraH218O} and show narrow ($\sim$3.5 km s$^{-1}$) line profiles, consistent with emission arising in the inner envelope. The blue side of the H$_2^{18}$O line overlaps with a CH$_3$OCH$_3$ line and HDO has some overlap with a weak CH$_3$OCHO line at the highest blueshifted velocities. These blended channels have been excluded in the integrated intensity maps, but in both cases, the blending line shows a similar spatial extent as the water line. The H$_2^{18}$O emission is marginally resolved and extends out to 200--300 au. The HDO observations have higher resolution (75$\times$48 au versus 280$\times$175 au), and the marginally resolved HDO emission extends out to $\sim$100 au. 

Deconvolving the moment zero maps using the CASA \textit{imfit} function results in an elliptical component with a major axis of 93 $\pm$ 58 au perpendicular to the outflow and a minor axis of 35 $\pm$ 44 au for H$_2^{18}$O and a more spherical component of 39 $\pm$ 8.7 $\times$ 38 $\pm$ 10.5 au for HDO. Assuming the H$_2^{18}$O and HDO emission arise from the same region, which is a reasonable assumption given their comparable upper level energy, the emitting area of water is better constrained by the higher resolution and higher sensitivity HDO observations. This is supported by the fact that the minor axis of the H$_2^{18}$O component is very similar to the HDO results, while the larger major axis of the Gaussian fit is along the major axis of the beam. Adopting the fitted semi-minor axis as estimate of the snowline radius then results in a snowline at 18 $\pm$ 22 au based on H$_2^{18}$O and at 19 $\pm$ 6 au based on HDO. 

Assuming the emission is optically thin and in local thermodynamic equilibrium (LTE), the H$_2^{18}$O and HDO column densities, $N_T$, can be calculated using 
\begin{equation}\label{eq1}
 \frac{4\pi F\Delta v}{A_{ul}\Omega hcg_{\mathrm{up}}} = \frac{N_T}{Q(T_{\mathrm{ex}})}e^{-E_{\mathrm{up}}/kT_{\mathrm{ex}}}, 
\end{equation}
where $F\Delta v$ is the integrated flux density, $A_{ul}$ is the Einstein A coefficient, $\Omega$ is the solid angle subtended by the source, $E_{\mathrm{up}}$ and $g_{\mathrm{up}}$ are the upper level energy and degeneracy, respectively, $T_{\mathrm{ex}}$ is the excitation temperature and $Q(T_{\mathrm{ex}})$ is the partition function. We adopt the molecular line parameters from the Jet Propulsion Laboraty (JPL) database \citep{Pickett1998}, where the submillimeter line measurements for H$_2^{18}$O are from \citet{DeLucia1972} and those for HDO are from \citet{Messer1984}. The H$_2^{18}$O line is a para transition, so we adopt the para-H$_2^{18}$O partition function and an ortho/para ratio of 3 to calculate the total H$_2^{18}$O column density. We assume an excitation temperature of 124~K, as adopted by previous studies of warm water in protostellar envelopes \citep{Persson2014,Jensen2019}. Increasing the temperature to 300 K increases the column densities by less than 40\%. 

\begin{figure*}
\centering
\includegraphics[width=\textwidth,trim={0cm 12.5cm 0cm 1.4cm},clip]{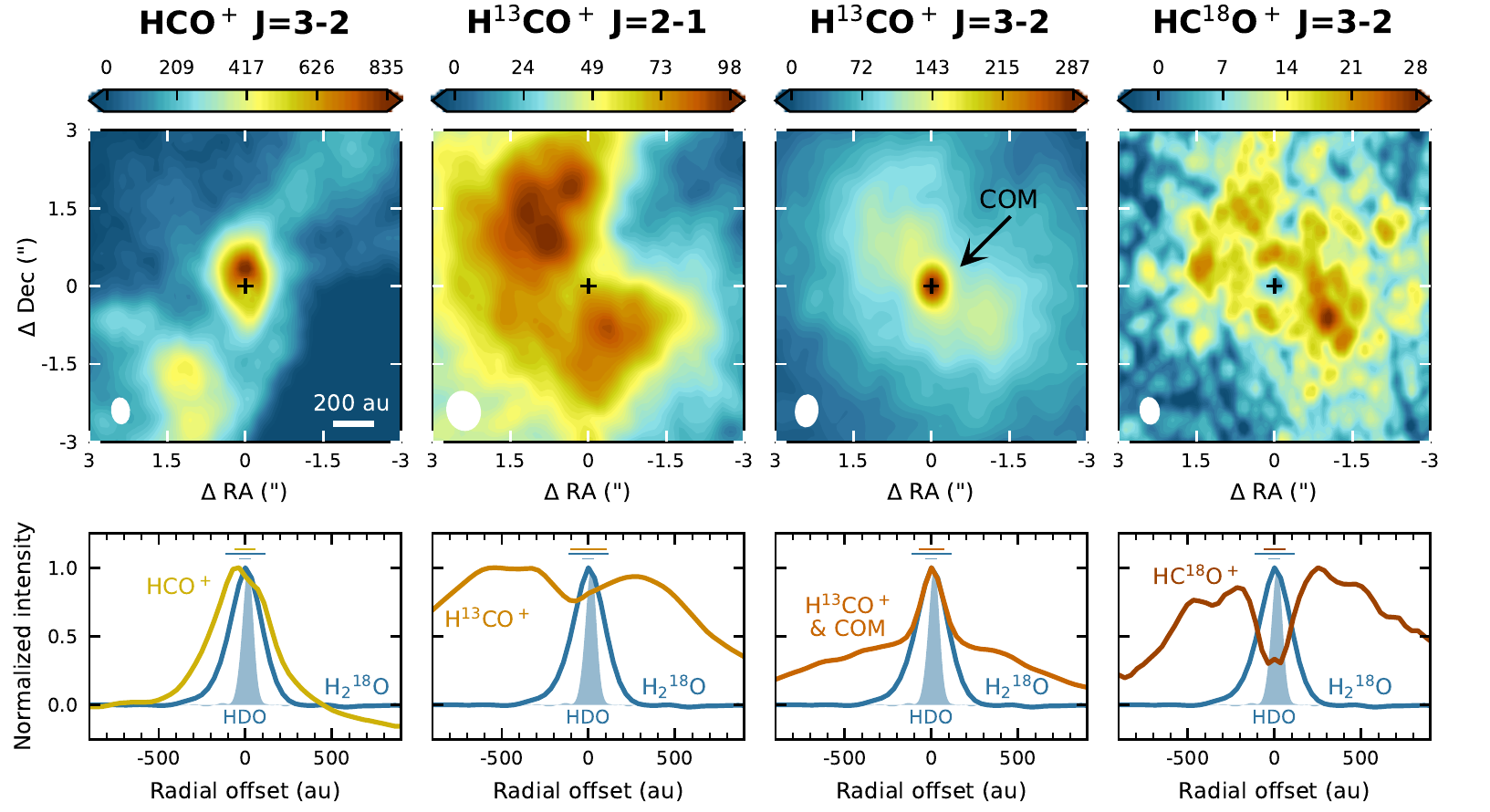}
\caption{Integrated intensity maps for the HCO$^+$ $J=3-2$, H$^{13}$CO$^+$ $J=2-1$, H$^{13}$CO$^+$ $J=3-2$ and HC$^{18}$O$^+$ $J=3-2$ transitions toward B1-c (top panels), and corresponding azimuthally averaged radial intensity profiles (bottom panels). H$^{13}$CO$^+$ $J=3-2$ is blended with a line from the complex organic molecule (COM) CH$_3$OCHO. Negative (positive) radial offsets correspond to the east (west). Averages are taken over position angles ranging from 0 to 90$^\circ$ to avoid the outflow cavity, except for HC$^{18}$O$^+$ $J=3-2$ for which the position angle range is taken 0--180$^\circ$. The azimuthally averaged radial intensity profiles for H$_2^{18}$O and HDO $3_{1,2}-2_{2,1}$ are shown using a blue line and blue-shaded area, respectively. The black cross in the top panels marks the continuum peak and the color scale is in mJy beam$^{-1}$ km s$^{-1}$. The beam size is depicted in the lower left corner of the top panels, and indicated by the horizontal lines in the top center of the bottom panels. }
\label{fig:B1-c}
\end{figure*}

From the \textit{imfit} procedure we obtain integrated fluxes of 126 $\pm$ 13 mJy km s$^{-1}$ and 241 $\pm$ 14 mJy km s$^{-1}$ for H$_2^{18}$O and HDO, respectively. This results in respective column densities of (1.3 $\pm$ 0.1) $\times 10^{17}$ cm$^{-2}$ and (5.5 $\pm$ 0.3) $\times 10^{16}$ cm$^{-2}$ for an emitting area 20 au in radius. Fluxes obtained from a Gaussian fit to the line profiles, to mitigate the effect of the water lines being partly blended, result in very similar values. These column densities are 5--50 times larger than previously found toward a small number of protostars (e.g., \citealt{Persson2012,Jensen2019}), but this is likely because the high resolution of the here presented HDO observations allowed us to better constrain the emitting area. Adopting an emitting area of 0.5--1.0$^{\prime\prime}$, similar to the beam sizes of earlier observations, results in column densities similar to those reported in earlier work. The column densities toward B1-c are $\sim$7 times smaller than derived toward the protostar IRAS 16293A where the emitting area was constrained by $\sim$0.3$^{\prime\prime}$ resolution observations \citep{Persson2013}. Using a $^{16}$O/$^{18}$O ratio of 560 \citep{Wilson1994} to determine the H$_2$O column density gives a HDO/H$_2$O ratio of (7.6 $\pm$ 0.9) $\times 10^{-4}$ toward B1-c, very similar to the ratios derived toward four other protostars in Perseus and Ophiuchus ($\sim 6-9 \times 10^{-4}$, \citealt{Persson2014,Jensen2019}). 

The H$_2^{18}$O observations also covered the much weaker HDO $3_{2,1}-4_{0,4}$ line at 207.110852 GHz (Einstein A coefficient of $1.1 \times 10^{-7}$ s$^{-1}$ versus $1.3 \times 10^{-5}$ s$^{-1}$). While emission is detected at this frequency (see Fig.~\ref{fig:SpectraH218O}), this is unlikely to be attributed to the HDO line, because the total flux of 50 $\pm$ 14 mJy km s$^{-1}$ suggests a column density of (2.1 $\pm$ 0.6) $\times 10^{18}$ cm$^{-2}$ and a HDO/H$_2$O ratio of $\sim$0.03. There are no other lines listed in Splatalogue\footnote{https://splatalogue.online}  close to the HDO frequency and the nearest line is a CH$_3$O$^{13}$CHO line $\sim$2 km s$^{-1}$ offset from the HDO frequency. Although such a large discrepancy between the two HDO lines from different datasets is very unlikely to be due to flux calibration errors, we compared the flux of the CH$_3$OH $12_5-13_4$ transition at 206.001302 GHz with the fluxes of the CH$_3$OH $5-4$ ladder around 241.8 GHz. The flux of the $12_5-13_4$ line is $\sim$60 mJy, and the $5-4$ lines are clearly optically thick with a flux of $\sim$80--85 mJy. At a column density of $2 \times 10^{18}$ cm$^{-2}$ and excitation temperatures of 100--200 K \citep{vanGelder2020}, the $12_5-13_4$ line will also be optically thick and will then have a flux of 57--61 mJy. These results thus suggest that there is not a large error in flux calibration, and the emission at 207.111852 GHz is most likely from an unidentified line or maybe high velocity outflow emission.

\subsection{H\textsuperscript{13}CO$^+$} \label{sec:B1-c_H13CO+}

Figure~\ref{fig:M0_H13COp-H218O} provides a textbook example of compact H$_2^{18}$O emission surrounded by a ring of H$^{13}$CO$^+$ ($J=2-1$) toward B1-c, as expected for HCO$^+$ destruction by water. The H$^{13}$CO$^+$ emission peaks $\sim$300 au ($\sim$1.0$^{\prime\prime}$) off source, with the ring shape disrupted along the direction of the outflow axes, especially for the redshifted outflow. A very similar morphology has been observed toward this source on larger scales for N$_2$H$^+$, which peaks outside the CO snowline \citep{Matthews2006}. Large-scale H$^{13}$CO$^+$ emission in the central velocity channels is resolved out and there is some redshifted absorption toward the continuum peak (see Fig.~\ref{fig:LineprofilesH13COp-21}). Channels with absorption are excluded in the integrated intensity (moment zero) map. Including these channels only lowers the emission on source, and does not alter the ring-shaped morphology. The central depression is not due to optically thick continuum because the brightness temperature of the continuum is only a few K, and several other lines with various upper level energies are peaking on source (see \citealt{vanGelder2020,Nazari2021}). 
The ring shape is also not due to continuum subtraction because the H$^{13}$CO$^+$ emission peaks at the same radius when imaged without continuum subtraction.  

A comparison with the higher resolution HDO observations, and the derived snowline location of $\sim$20 au, clearly shows that the H$^{13}$CO$^+$ peak ($\sim$300 au) is only providing an upper limit to the snowline location. This is consistent with observations toward NGC1333 IRAS2A \citep{vantHoff2018a}. A better constraint may be obtained from higher resolution H$^{13}$CO$^+$ observations as the relatively large beam could spread out the signature of a steep rise in column density. However, chemical modeling shows that there is an offset between the HCO$^+$ column density peak and the snowline \citep{vantHoff2018a,Hsieh2019,Leemker2021}, and the radius of the HCO$^+$ emission peak is also influenced by whether a disk is present in the innermost region as well as the source inclination angle \citep{Hsieh2019}. Therefore, deriving a more stringent snowline location from H$^{13}$CO$^+$ emission requires radiative transfer modeling using a source specific physical structure. 

\begin{figure*}
\centering
\includegraphics[trim={0.5cm 15.6cm 0cm 1.5cm},clip]{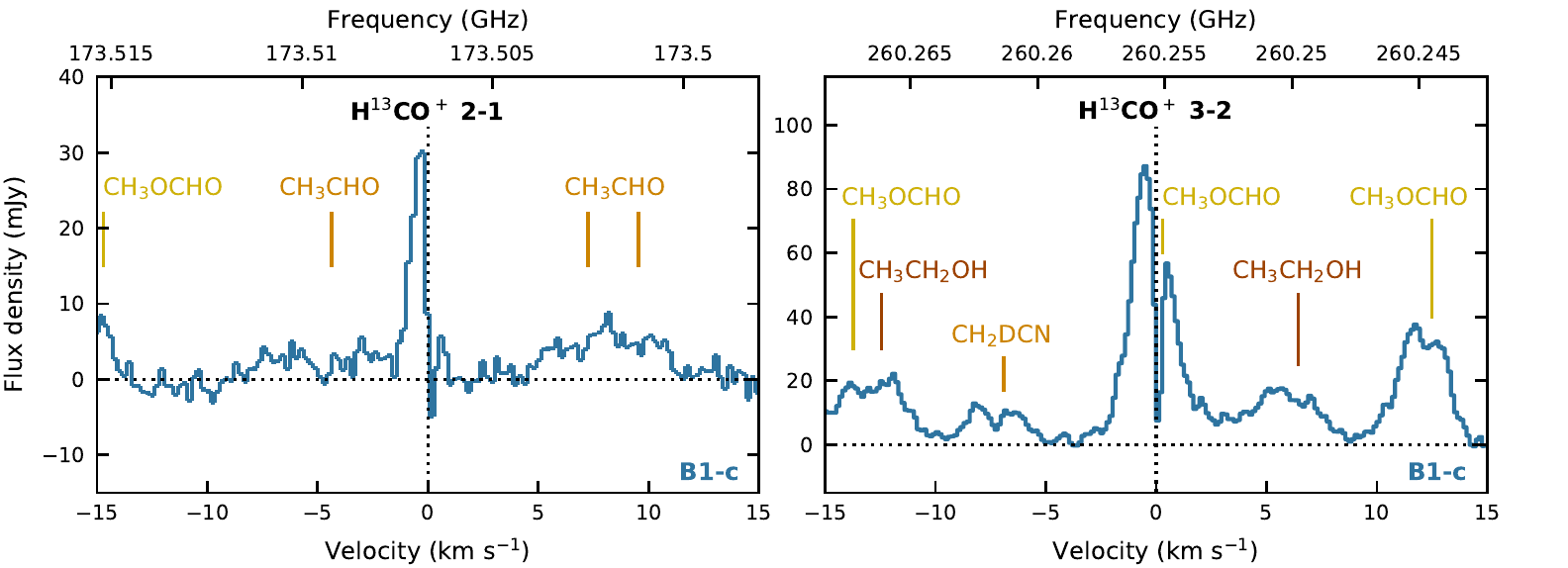}
\caption{Spectra extracted within a circular 0.5$^{\prime\prime}$ diameter aperture ($\sim$one beam) toward the continuum peak position of B1-c centered around the H$^{13}$CO$^+$ $J=2-1$ (left panel) and $J=3-2$ (right panel) transitions. The $J=3-2$ transition is blended with a CH$_3$OCHO line, while the $J=2-1$ transition is unblended.}
\label{fig:Spectra_H13COp_21-32}
\end{figure*}

In addition to the $J=2-1$ transition, H$^{13}$CO$^+$ has been observed toward B1-c using the $J=3-2$ transition at slightly better resolution ($0.6^{\prime\prime} \times 0.4^{\prime\prime}$ versus $0.7^{\prime\prime} \times 0.6^{\prime\prime}$). Moment zero maps and radial profiles for both transitions are presented in Fig.~\ref{fig:B1-c}. In contrast to the ring-shaped morphology of the $J=2-1$ line, the $J=3-2$ line displays a bright central component surrounded by weaker extended emission. The bright unresolved component is due to a methyl formate (CH$_3$OCHO) line 259 kHz (0.3 km s$^{-1}$) offset from the H$^{13}$CO$^+$ $J=3-2$ line (Fig.~\ref{fig:Spectra_H13COp_21-32}). This CH$_3$OCHO line at 260.25508 GHz should be as strong as the line at 260.24450 GHz because they form a doublet. As can be seen in Fig.~\ref{fig:Spectra_H13COp_21-32}, this second line is clearly detected. In addition, \citet{vanGelder2020} detected 12 CH$_3$OCHO lines in the H$^{13}$CO$^+$ $J=3-2$ dataset and both lines discussed here are consistent with CH$_3$OCHO emission from a $2\times10^{16}$ cm$^{-2}$ column with an excitation temperature of 180 K. The H$^{13}$CO$^+$ $J=3-2$ line is thus less suited to trace the water snowline than the $J=2-1$ line for which no transitions from complex organics are listed in Splatalogue within 1 MHz (see Fig.~\ref{fig:Spectra_H13COp_21-32}). 

The H$^{13}$CO$^+$ $J=4-3$ transition has not been observed toward B1-c, but its rest frequency of 346.99834 GHz is very close to that of acetaldehyde (CH$_3$CHO) lines at 346.99553 and 346.99955 GHz. Observations show that H$^{13}$CO$^+$ $J=4-3$ is indeed blended with acetaldehyde lines in the disk around the outbursting young star V883-Ori, making it harder to trace the water snowline \citep{Lee2019,Leemker2021}. The upper level energy of 4~K makes the H$^{13}$CO$^+$ $J=1-0$ transition more sensitive to colder extended material, and the flux from the warm inner envelope will be weak compared to the flux from higher excitation lines. For H$^{13}$CO$^+$, the $J=2-1$ transition is therefore the best line to trace the water snowline in line-rich sources, while stronger higher energy transitions may possibly be used in sources that lack emission from complex organics. 

\subsection{HCO$^+$ and HC\textsuperscript{18}O$^+$}\label{sec:B1-c_isotopologues}

Figure~\ref{fig:B1-c} also presents moment zero maps and radial profiles for the HCO$^+$ $J=3-2$ and HC$^{18}$O$^+$ $J=3-2$ transitions toward B1-c. The HCO$^+$ data have been presented previously by \citet{Hsieh2019}, and as done in that work, we exclude the central channels that display absorption. HC$^{18}$O$^+$ $J=3-2$ displays a similar ring-shaped morphology as H$^{13}$CO$^+$ $J=2-1$ (see also Fig.~\ref{fig:M0_H13COp-H218O}), but the emission peaks slightly closer to the protostar (200--250 au). This could be a result of the higher spatial resolution of the HC$^{18}$O$^+$ data. Most interestingly, the HDO emission falls within the central depression ($\sim$40 au radius) visible in the HC$^{18}$O$^+$ image, suggesting that the inner radius of the emission is a better tracer of the snowline than the emission peak. This cavity and inner radius is likely less defined in the H$^{13}$CO$^+$ $J=2-1$ image due to a larger contribution from the outer envelope for this more abundant isotopogue. The cavity may be more visible in the $J=3-2$ transition, which is more sensitive to warmer and denser material, but we cannot confirm this due to the blending with a methyl formate line. In contrast, the HCO$^+$ emission shows a compact component that peaks $\sim$75 au off source, and more extended emission along the outflow directions. 

H$^{13}$CO$^+$ and HC$^{18}$O$^+$ are expected to be better tracers of the water snowline than the main isotopologue HCO$^+$, because they are less abundant and therefore less optically thick or even optically thin. As such, they will trace the higher density inner region of the envelope. \citet{Jorgensen2009} estimated that HCO$^+$ $J=3-2$ emission does not trace the inner 100 au, due to the emission becoming optically thick in the outer envelope, for envelope masses $>0.1M_\odot$. With an envelope mass of 3.8 $M_\odot$ \citep{Enoch2009}, the HCO$^+$ $J=3-2$ emission is thus expected to be optically thick in B1-c. This assumption can now be tested observationally by comparing the HCO$^+$ emission to the emission from the less abundant isotopologues H$^{13}$CO$^+$ and HC$^{18}$O$^+$. 

For optically thin emission, the HCO$^+$/H$^{13}$CO$^+$ and HCO$^+$/HC$^{18}$O$^+$ ratios should be similar to the elemental [$^{12}$C]/[$^{13}$C] ratio of 77 and [$^{16}$O]/[$^{18}$O] ratio of 560, respectively \citep{Wilson1994}. At radii larger than 300 au, that is, to avoid the methyl formate contamination to the H$^{13}$CO$^+$ $J=3-2$ emission, the HCO$^+$ ($J=3-2$) / H$^{13}$CO$^+$ ($J=3-2$) line ratio is $\sim$2--3. The HCO$^+$ ($J=3-2$) / HC$^{18}$O$^+$ ($J=3-2$) line ratio is $\sim$200 on source, where the HC$^{18}$O$^+$ emission is low, and decreases to $\lesssim$50 in the HC$^{18}$O$^+$ ring at $\gtrsim$200 au. The H$^{13}$CO$^+$ ($J=3-2$) / HC$^{18}$O$^+$ ($J=3-2$) line ratio is $\sim$ 5--7 outside of the central gap. These ratios suggests that the HCO$^+$ emission is optically thick, while the H$^{13}$CO$^+$ and HC$^{18}$O$^+$ emission are optically thin, at least at radii $\gtrsim$300 au. For temperatures between 40--80 K, the H$^{13}$CO$^+$ $J=3-2$ transition is expected to be $2-3$ times as bright as the $J=2-1$ line based on radiative transfer calculations with RADEX \citep{vanderTak2007}, as long as the emission is optically thin. The observed H$^{13}$CO$^+$ $J=3-2$/$J=2-1$ ratio is $\sim$2.5 at radii $\gtrsim$300 au, consistent with optically thin emission at 50--60 K. The H$^{13}$CO$^+$ optical depth may be higher in the central velocity channels toward the continuum peak as evidenced by a slight absorption feature for the 2--1 transition (Fig.~\ref{fig:LineprofilesH13COp-21}), but these channels are excluded in the moment zero map. When imaged before continuum subtraction, the H$^{13}$CO$^+$ $J=2-1$ emission peaks at the same radius as displayed in Figs.~\ref{fig:M0_H13COp-H218O} and \ref{fig:B1-c} (after continuum subtraction). This shows that the central depression is not due to the subtraction of continuum emission from optically thick line emission. 

The isotopologue emission is thus likely optically thin, and follows the column density profile, in contrast to the main isotopologue HCO$^+$. As discussed in more detail by \citet{Hsieh2019}, line self-absorption and/or continuum subtraction can create a central hole that is unrelated to the snowline if the HCO$^+$ is optically thick. Considering the emission of all isotopologues, these effects are thus likely the reason for the small dip in HCO$^+$ emission, while H$^{13}$CO$^+$ and HC$^{18}$O$^+$ trace the column density decrease inside the snowline. These observations thus show that that H$^{13}$CO$^+$ is indeed a better tracer of the water snowline than its main isotopologue. Moreover, HC$^{18}$O$^+$ is an even better tracer due to the smaller contribution from the outer envelope and lower optical depth. 

\begin{figure}
	\centering
	\subfloat{\includegraphics[trim={0.5cm 14.2cm 9cm 0.7cm},clip]{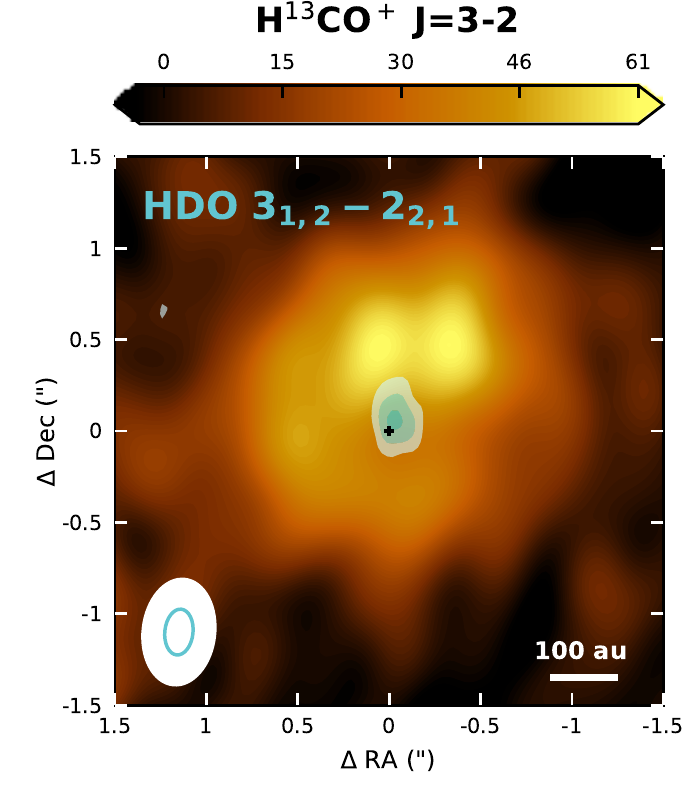}}
	\hfill
	\subfloat{\includegraphics[trim={0.5cm 16.3cm 9cm 2.1cm},clip]{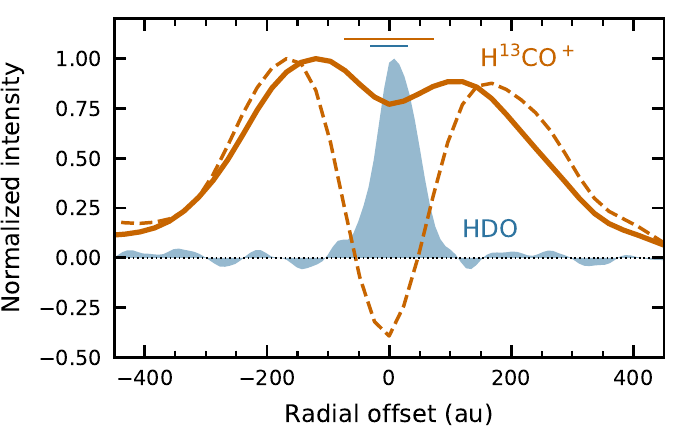}}
	\caption{Top panel: Integrated intensity map for the H$^{13}$CO$^+$ $J=3-2$ transition (black to yellow color scale in mJy beam$^{-1}$ km s$^{-1}$) toward B1-bS. Channels containing absorption are not included. The integrated intensity map for the HDO $3_{1,2}-2_{2,1}$ transition is overlaid in transparent shades of blue. The HDO image is clipped at $3\sigma$ and the contours are at [3, 6, 9]$\sigma$, where $\sigma = 1.7$ mJy beam $^{-1}$ km s$^{-1}$. The continuum peak is marked by a cross and the beams are depicted in the lower left corners (solid white for H$^{13}$CO$^+$ and blue contour for HDO). Bottom panel: Corresponding azimuthally averaged radial intensity profiles for H$^{13}$CO$^+$ (solid orange line) and HDO (blue-shaded area). The dashed orange line shows the H$^{13}$CO$^+$ radial profile when channels containing absorption are included. The beam sizes are indicated by the horizontal lines in the top center. }
	\label{fig:B1-bS}
\end{figure}


\section{Imaging the water snowline in other protostellar envelopes} \label{sec:Envelopes}

\begin{figure*}
\centering
\includegraphics[width=\textwidth,trim={0cm 12.5cm 0cm 1.5cm},clip]{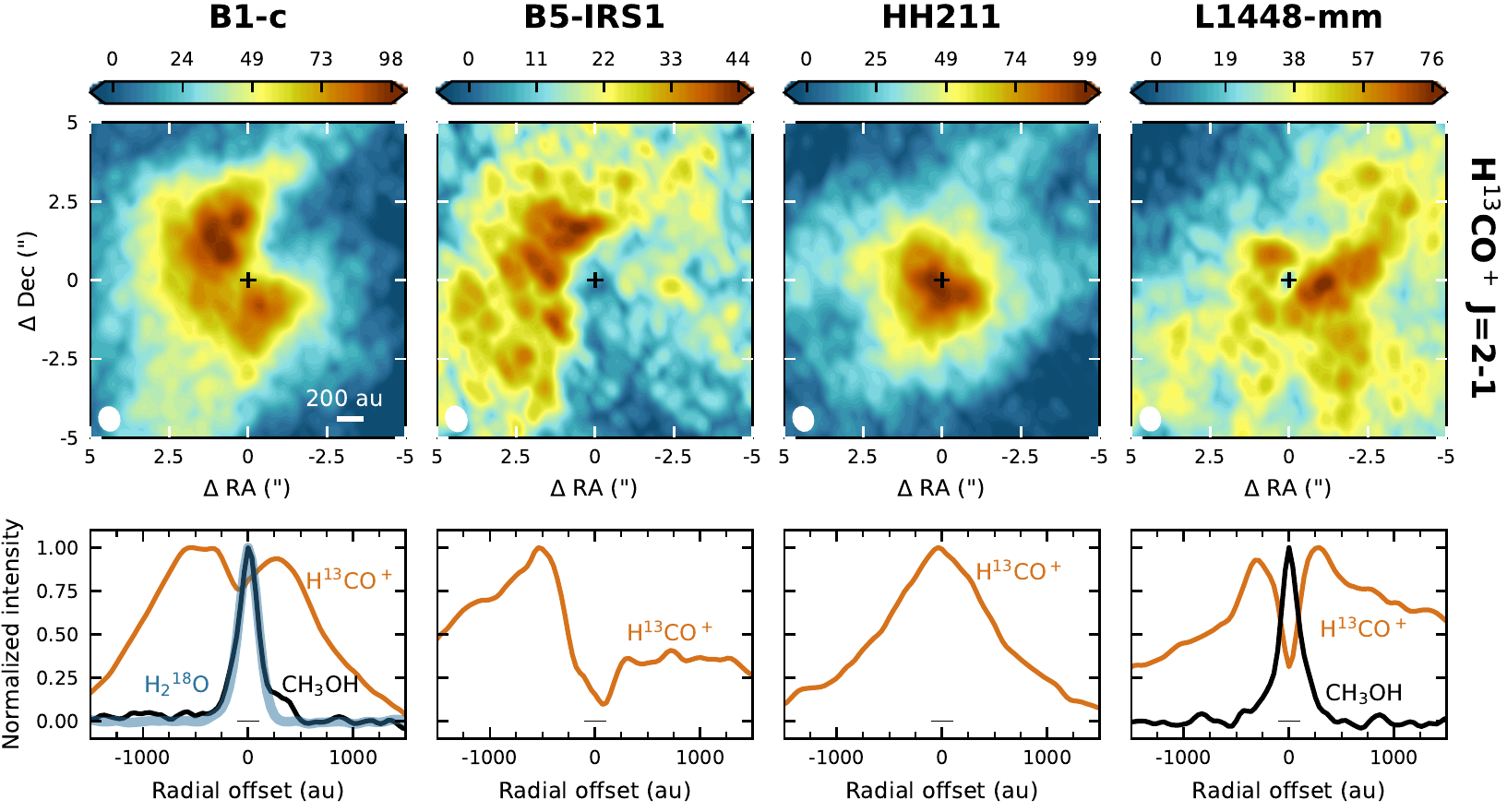}
\caption{Integrated intensity maps for the H$^{13}$CO$^+$ $J=2-1$ transition (top panels), and corresponding azimuthally averaged radial intensity profiles (orange line, bottom panels). For B1-c and L1448-mm, H$^{13}$CO$^+$ channels with central absorption are excluded (see Fig.~\ref{fig:LineprofilesH13COp-21}). For B1-c and L1448-mm, the radial intensity profile of the CH$_3$OH $15_{1,15}-14_{ 2,12}$ transition at 187.5429 GHz is shown in black, and for B1-c the H$_2^{18}$O profile is shown in blue. Negative (positive) radial offsets correspond to the east (west). Averages are taken over position angles ranging from 0 to 180$^\circ$, except for H$^{13}$CO$^+$ toward B1-c and B5-IRS1. For B1-c, the range was limited to 0--90$^\circ$ to avoid the outflow cavity and for B5-IRS1 the range was limited to 0--160$^\circ$ to follow the displayed arc shape. The black cross in the top panels marks the continuum peak and the color scale is in mJy beam$^{-1}$ km s$^{-1}$. The beam size is depicted in the lower left corner of the top panels, and indicated by the horizontal lines in the bottom center of the bottom panels.}
\label{fig:H13COp-21}
\end{figure*}

The most direct measurement of the snowline comes from water observations, and B1-bS is the only other source in our sample for which water (HDO) has been observed and detected (Figs.~\ref{fig:SpectraH218O} and \ref{fig:B1-bS}). H$_2^{18}$O was observed, but not detected toward HH211 (Fig.~\ref{fig:SpectraH218O}), and HDO was observed but not detected toward B1-bN. B1-bN was not covered by any of the other observing programs, so we discuss the HDO upper limit in Appendix~\ref{app:B1-bN}. H$^{13}$CO$^+$ $J=2-1$ has been observed toward B5-IRS1, HH211 and L1448-mm (Fig.~\ref{fig:H13COp-21}), and H$^{13}$CO$^+$ $J=3-2$ toward B1-bS (Fig.~\ref{fig:B1-bS}). The H$^{13}$CO$^+$ lines are narrow ($\sim$1~km~s$^{-1}$; see Figs.~\ref{fig:LineprofilesH13COp-21} and \ref{fig:Spectra_B1-bS}), indicating that the emission arises from the inner envelope and not the outflow. In addition to B1-c, ring-shaped H$^{13}$CO$^+$ emission is observed toward B1-bS, surrounding the HDO emission, and L1448-mm. B5-IRS1 displays and arc of H$^{13}$CO$^+$ emission northeast of the continuum peak, while the emission peaks on source in HH211. We will discuss the sources individually in the next sections (Sect.~\ref{sec:B1-bS}-\ref{sec:B5-IRS1}). The full spatial extent of the H$^{13}$CO$^+$ emission is shown in Fig.~\ref{fig:largescale}. 

\subsection{B1-bS} \label{sec:B1-bS}

The moment zero map and radial profile of the HDO $3_{1,2}-2_{2,1}$ transition toward B1-bS are presented in Fig.~\ref{fig:B1-bS}. The HDO emission is centrally peaked and marginally resolved. Deconvolution with the CASA task \textit{imfit} returns a major axis of 56 $\pm$ 33.3 au and a minor axis of 40 $\pm$ 17.7 au. The major axis is roughly along the major axis of the beam, and under the assumption that HDO emits from a roughly spherical region, we adopt half of the deconvolved minor axis as an estimate of the snowline radius. This then gives a snowline radius of 20 $\pm$ 9 au. The HDO emission is more compact than the continuum, which has a deconvolved size of 133 $\pm$ 1.4 au $\times$ 122 $\pm$ 1.5 au. The total flux of 31 $\pm$ 8.7 mJy km s$^{-1}$ results in a HDO column density of $(7.0 \pm 2.0) \times 10^{15}$ cm$^{-2}$, about an order of magnitude lower than toward B1-c.  

The H$^{13}$CO$^+$ $J=3-2$ emission displays a ring-shaped morphology, surrounding the HDO emission and peaking around 125 au (Fig.~\ref{fig:B1-bS}). Methyl formate emission is detected, just as for B1-c, but the blending may be less pronounced due to the strong redshifted absorption of the H$^{13}$CO$^+$ line (see Fig.~\ref{fig:Spectra_B1-bS}). Channels with absorption are excluded from the moment zero map, and including them shifts the peak of the H$^{13}$CO$^+$ emission $\sim$50 au outward (see radial profiles in Fig.~\ref{fig:B1-bS}). Translating a H$^{13}$CO$^+$ emission peak into a snowline location thus gets more complicated in sources with a strong envelope contribution displayed as redshifted absorption features.  

\subsection{L1448-mm} \label{sec:L1448-mm}

As for B1-c, a ring-like morphology is observed for H$^{13}$CO$^+$ toward L1448-mm, with the emission peaking $\sim$300 au off source. The overall distribution is more asymmetric with respect to the outflow axis in L1448-mm, with the redshifted emission in the southwest extending out to larger radii than the blueshifted emission in the northeast. This is likely due to the observed blueshifted absorption (see Fig.~\ref{fig:LineprofilesH13COp-21}), which may be caused by a wide-opening angle wind \citep[e.g.,][]{Hirano2010}. Including the channels with blueshifted absorption in the moment zero map makes the emission peak in the northeast roughly as bright as the peak in the southwest, but does not affect the overall emission morphology. 

A few lines from complex organics are detected toward L1448-mm, and although different species have different freeze-out temperatures, emission from lines with upper-level energies $\gtrsim$100 K roughly originate inside the water snowline. The radial profile of the CH$_3$OH $15_{1,15}-14_{2,12}$ transition ($E_{\rm{up}}$ = 290 K) at 187.5429 GHz is compared to the H$^{13}$CO$^+$ profile for B1-c and L1448-mm in Fig.~\ref{fig:H13COp-21}, and an overlay of the moment zero maps is shown in Fig.~\ref{fig:CH3OH}. The spatial extent of the CH$_3$OH emission is similar in both sources, and for B1-c this is similar to the spatial extent of H$_2^{18}$O. The similarity between both the H$^{13}$CO$^+$ and the CH$_3$OH morphology toward B1-c and L1448-mm suggest a similar snowline location in these sources. 

A Gaussian fit in the image plane to the moment zero map of the methanol emission results in a snowline estimate of 123 $\pm$ 71 au for B1-c and 100 $\pm$ 50 au for L1448-mm. This estimate for B1-c is larger than derived from H$_2^{18}$O and HDO (18 $\pm$ 22 au and 19 $\pm$ 6 au, respectively). This is likely due to the lower signal-to-noise of the methanol observations as well as more extended methanol emission at the $\sim$2$\sigma$ level. This $\sim$2$\sigma$ extended emission is also observed toward L1448-mm. We therefore use the similarity in H$^{13}$CO$^+$ and CH$_3$OH between B1-c and L1448-mm to tentatively estimate a snowline radius of $\sim$20 au in L1448-mm. 

\subsection{HH211} \label{sec:HH211}

Compact, centrally peaked H$^{13}$CO$^+$ emission is observed toward HH211 (Fig.~\ref{fig:H13COp-21}), while H$_2^{18}$O is not detected (Fig.~\ref{fig:SpectraH218O}). The H$_2^{18}$O non-detection is consistent with the absence of emission lines from complex organic molecules. When imaged at slightly higher resolution using robust weighting of 0.5 ($0.46^{\prime\prime} \times 0.62^{\prime\prime}$) there is a tentative depression ($\sim$2--3$\sigma$) toward the source position with the H$^{13}$CO$^+$ emission peaking around 75 au. If the dip is real, this would suggest that the snowline is located closer in than 75 au. 

A 3$\sigma$ upper limit for the H$_2^{18}$O column density can be calculated by substituting 
\begin{equation} \label{eq2}
3\sigma = 3 \times1.1 \sqrt{\delta v\Delta V} \times \mathrm{rms}, 
\end{equation}
for the integrated flux density, $F\Delta v$ in Eq.~\ref{eq1}. Here $\delta v$ is the velocity resolution, and $\Delta V$ is the line width which we take to be the same as observed for B1-c ($\sim$3.5 km s$^{-1}$). The factor 1.1 takes a 10\% calibration uncertainty in account. The rms in the spectrum extracted in the central beam amounts to 2.2 mJy, resulting in a column density upper limit of 4.4 $\times 10^{13}$ cm$^{-2}$ assuming the emission fills the beam. A more narrow line width of 1 km s$^{-1}$ would reduce the column by only a factor $\sim$2. If the snowline would be at 20 au, as for B1-c, the column density upper limit would be 4.2 $\times 10^{15}$ cm$^{-2}$, 30 times lower than toward B1-c. A snowline radius of 3.5 au would result in a upper limit similar to the column in B1-c. This thus suggests that either the water column, and possibly the abundance, toward HH211 is low, or the snowline is only a few au from the star. 

\subsection{B5-IRS1}\label{sec:B5-IRS1}

The H$^{13}$CO$^+$ emission toward B5-IRS1 seems to originate predominantly in a ridge-like structure peaking $\sim$ 500 au northeast of the source that extends out to larger scales in the northwest and southeast than displayed in Fig.~\ref{fig:H13COp-21} (see Fig.~\ref{fig:largescale}). The peak of the emission toward this target is $\sim$2 times weaker compared to the other sources. The only other molecules we detected are large scale emission from H$^{13}$CN, H$_2$CS and D$_2$CO, that is, these lines are only detected in spectra integrating over a $\gtrsim$5$^{\prime\prime}$ diameter aperture. If the H$^{13}$CO$^+$ emission is associated with the snowline, it suggests a larger snowline radius than toward B1-c ($<$ 500 au based on the H$^{13}$CO$^+$ peak), and lower column densities of complex organics inside the snowline. The arc-like structure of H$^{13}$CO$^+$ could also mean that the emission is associated with larger scales in B5-IRS1 rather than the inner envelope. Water observations are required to solve this degeneracy.


\section{The water snowline and protostellar accrection bursts} \label{sec:Bursts}

\begin{deluxetable*}{lcccccccccccc}
\tablecaption{Predicted and observed water snowline locations. \label{tab:Snowline}}
\tablewidth{700pt}
\addtolength{\tabcolsep}{-2pt} 
\tabletypesize{\scriptsize}
\tablehead
{
\colhead{\vspace{-3mm}}&
\multicolumn{3}{c}{Current luminosity}&
\multicolumn{2}{c}{Burst luminosity\tablenotemark{a}}&
\multicolumn{2}{c}{Burst luminosity\tablenotemark{b}}&
\multicolumn{2}{c}{Burst luminosity\tablenotemark{c}}&
\multicolumn{3}{c}{This work}\\
\colhead{}&
\multicolumn{3}{c}{}&
\multicolumn{2}{c}{(J{\o}rgensen}&
\multicolumn{2}{c}{(Frimann}&
\multicolumn{2}{c}{(Hsieh}&
\multicolumn{3}{c}{}\\
\colhead{}&
\multicolumn{3}{c}{}&
\multicolumn{2}{c}{et al. 2015)}&
\multicolumn{2}{c}{et al. 2017)}&
\multicolumn{2}{c}{et al. 2019)}&
\multicolumn{3}{c}{}\\
\cline{2-4} \cline{5-6} \cline{7-8} \cline{9-10} \cline{11-13}
\colhead{Source \vspace{-3mm}}& 
\colhead{$L$}& \colhead{ref\tablenotemark{d}} & \colhead{$R_{\rm{snow}}$\tablenotemark{e}}&  
\colhead{$L_{\rm{burst}}$}& \colhead{$R_{\rm{snow}}$\tablenotemark{e}}&  
\colhead{$L_{\rm{burst}}$}& \colhead{$R_{\rm{snow}}$\tablenotemark{e}}& 
\colhead{$L_{\rm{burst}}$}& \colhead{$R_{\rm{snow}}$\tablenotemark{e}}& 
\colhead{$R_{\rm{snow}}$\tablenotemark{f}}& \colhead{$L_{\rm{snow}}$\tablenotemark{g}} & \colhead{Last burst}\\
\colhead{} & 
\colhead{($L_\odot$)}& \colhead{}& \colhead{(au)}&  
\colhead{($L_\odot$)}& \colhead{(au)}&  
\colhead{($L_\odot$)}& \colhead{(au)}&  
\colhead{($L_\odot$)}& \colhead{(au)}& 
\colhead{(au)}& \colhead{($L_\odot$)} & \colhead{(yr)}
}
\startdata
B1-c 				& 5.2 		& 1 		& 35 	& - 		& -		& 13--67 	& 55--126		& 9.4--47 	& 47--106 	& 19$\pm$6 -- 20$\pm$9			& 1.5$\pm$0.9 -- 1.7$\pm$1.5		& $>$ 10,000 			\\
B1-bS			& 0.5 		& 2 		& 11		& - 		& -		& -			& -				& - 				& -				& 20$\pm$9								& 1.7$\pm$1.5 								& $>$ 1,000 				\\
B5-IRS1      	& 7.7 		& 3 		& 43		& - 		& -		& 15--68 	& 59--127 	& - 				& - 				& $<$ 500			 					& $<$ 1054								& $<$ 10,000	\\
HH211         	& 3.6 		& 1 		& 29 	& - 		& -		& 9.6--47	& 48--106 	& - 				& - 				& $<$ 20									& $<$ 1.7										& 1,000--10,000 	\\
L1448-mm 	& 9.0       	& 1 		& 46 	& - 		& -		& 10--51 	& 49--109 	& - 				& - 				& $\sim$ 20 $\blacklozenge$ 	& $\sim$ 1.7									& 1,000--10,000	\\\hline
B335             & 2.0       	& 4 		& 22		& 9.5	& 48 	& -			& -				& - 				& - 				& 10$\pm$5 -- 14$\pm$5 		& 0.4$\pm$0.4 -- 0.8$\pm$0.6 	& $>$ 10,000			\\
BHR71-IRS1  & 15.0      	& 5 		& 60		& - 		& -		& -			& -				& - 				& - 				& 44$\pm$ 3				 			& 8.2$\pm$4.8							& $>$ 1,000				\\
L483             & 10--13  	& 6,7	& 52		& - 		& -		& -			& -				& - 				& - 				& 14$\pm$12 -- 22$\pm$5 $\blacklozenge$	& 0.8$\pm$1.5 -- 2.0$\pm$0.9 		& $>$ 1,000				\\
IRAS 15398  & 1.8 		& 8		& 21		& - 		& -		& -			& -				& - 				& - 				& $\sim$ 100 $\bigstar$			& $\sim$ 42								& $<$ 1,000				\\
IRAS 16293A & 18.0 		& 9 		& 65 	& - 		& -		& -			& -				& - 				& - 				& 66$\pm$24 -- 102$\pm$42	& 18$\pm$13 -- 24$\pm$9.5		& $>$ 1,000				\\
IRAS 2A         & 60.0 		& 1 		& 119	& 26		& 79 		& 10--49	& 49--108		& 26--104		& 78--157		& 60$\pm$15 -- 75$\pm$15		& 15$\pm$7.6 -- 24$\pm$9.5 		& 0 						\\
IRAS 4A-NW\tablenotemark{h} & 14.8 & 1 & 59     & 74 & 133 &  23--114 & 74--164 & - 		& - 				& 92$\pm$18							& 36$\pm$14 								& 1,000--10,000	\\
IRAS 4B         & 7.5 		& 1		& 42		& 10 		& 49		& 12--59 	& 54--118		& - 				& - 				& 30$\pm$15 							& 3.8$\pm$3.8 							& 1,000--10,000	\\
\enddata
\tablecomments{For sources in Perseus, luminosities are converted to a distance of 300 pc \citep{Ortiz-Leon2018}. For B335, the luminosities are converted to a distance of 165 pc \citep{Watson2020}.}
\vspace{-0.3cm}
\tablenotetext{a}{Burst luminosity determined from spatial extent of C$^{18}$O emission by \citet{Jorgensen2015}.}
\vspace{-0.3cm}
\tablenotetext{b}{Burst luminosities determined from spatial extent of C$^{18}$O emission by \citet{Frimann2017}. The lower value assumes a CO freeze-out temperature of 21 K, the higher value a CO freeze-out temperature of 28 K.}
\vspace{-0.3cm}
\tablenotetext{c}{Burst luminosity determined from CO snowline and H$_2$O snowline locations, which are derived from modeling N$_2$H$^+$ and HCO$^+$ emission, respectively \citep{Hsieh2019}.}
\vspace{-0.3cm}
\tablenotetext{d}{Luminosity references. [1] \citet{Karska2018}, [2] \citet{Hirano2014}, [3] \citet{Tobin2016}, [4] \citet{Evans2015}, [5] \citet{Tobin2019}, [6] \citet{Shirley2000}, [7] \citet{Tafalla2000}, [8] \citet{Jorgensen2013} [9] \citet{Jacobsen2018}.}
\vspace{-0.3cm}
\tablenotetext{e}{Snowline radius calculated using Eq.~\ref{eq:B07} \citep{Bisschop2007}.}
\vspace{-0.3cm}
\tablenotetext{f}{For the sources in the top part of the table, the snowline location is derived from H$^{13}$CO$^+$ and/or H$_2^{18}$O and HDO observations presented in this work. See Sections.~\ref{sec:B1-c_water} and \ref{sec:Envelopes} for details. For the sources in the bottom part of the table, snowline locations have been taken from the literature \citep{Persson2012,Persson2014,Jorgensen2013,Bjerkeli2016}, or derived here from observations presented before \citep{Jensen2019,Jensen2021}. See Appendix~\ref{app:literaturewater} for more details. A range in reported snowline radius reflects measurements using different water isotopologues. Snowline locations (excluding the upper limits for B5-IRS1 and HH211) consistent with a luminosity $> 5 L_{\rm{current}}$ are marked by a star and snowline locations corresponding to $< 1/5$ of $L_{\rm{current}}$ are marked by a black diamond.}
\vspace{-0.3cm}
\tablenotetext{g}{Luminosity corresponding to the derived snowline location listed in the preceding column calculated using Eq.~\ref{eq:B07} \citep{Bisschop2007}.}
\vspace{-0.3cm}
\tablenotetext{h}{Luminosities are for the IRAS4A binary.}
\end{deluxetable*}

\begin{figure*}
\centering
\includegraphics[trim={0cm 14.2cm 0cm 2.0cm},clip]{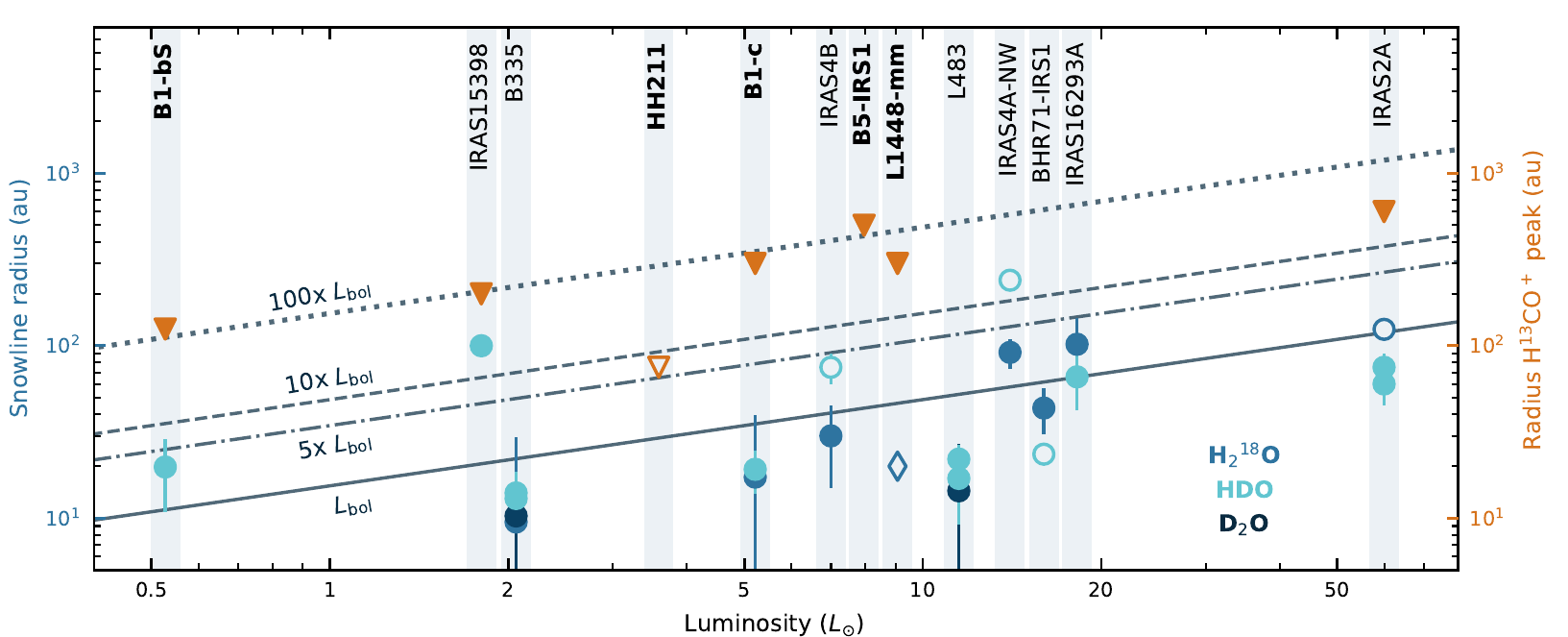}
\caption{Overview of the constraints on the water snowline location in the protostellar envelopes based on H$^{13}$CO$^+$ and/or water (H$_2^{18}$O, HDO or D$_2$O) observations. The peak of the H$^{13}$CO$^+$ emission ($J=4-3$ for IRAS 15398, $J=3-2$ for B1-bS and NGC1333-IRAS2A, and $J=2-1$ for the other sources), which serves as an upper limit of the snowline location, is shown as orange triangles. The tentative peak for HH211 is indicated by an open triangle. The snowline location derived from water observations is marked with a circle (blue for H$_2^{18}$O, light blue for HDO and dark blue for D$_2$O). Open circles indicate the presence of an outflow contribution. The snowline location derived for L1448-mm based on similarities in emission morphology with B1-c is marked with an open blue diamond. The expected snowline location based on bolometric luminosity (Eq.~\ref{eq:B07}) is indicated by the solid line. The dashed-dotted, dashed and dotted lines indicate the expected snowline locations if the source has recently undergone an accretion burst that increased the luminosity by a factor 5, 10 or 100, respectively. A source is considered to have recently undergone a burst if the snowline corresponds to a luminosity $>5 \times L_{\rm{current}}$. Names of sources studied in this work are highlighted in bold face, and Table~\ref{tab:LiteratureWater} lists references for sources not studied here. }
\label{fig:Snowlinelocations}
\end{figure*}

Without fully modeling the physical and chemical structure of a source, an estimate of the water snowline radius can be made based on the luminosity. \citet{Bisschop2007} derived the following relation from 1D radiative transfer modeling of high mass sources: 
\begin{equation}\label{eq:B07}
R_{\rm{snow}} \sim R(\rm{100 \hspace{1mm} K}) = 15.4 \sqrt{L/L_\odot} \hspace{2mm} \rm{au}.
\end{equation}
Radiative transfer modeling of low mass protostars shows that this relation also holds in the low-mass regime (see Appendix~\ref{app:high-low} and Fig.~\ref{fig:high-low}). We find an intrinsic uncertainty on the exact location of the snowline of 20--30$\%$ due to uncertainties in envelope properties (that is, the envelope mass and the density profile power-law index). This uncertainty corresponds to a few au for solar luminosity stars, and $\sim$10 au for 10 $L_{\odot}$ stars. Table~\ref{tab:Snowline} lists the snowline locations for the sources in our sample expected based on their luminosity using Eq.~\ref{eq:B07}, together with snowline locations derived from H$_2^{18}$O or H$^{13}$CO$^+$ in Sects.~{\ref{sec:B1-c} and \ref{sec:Envelopes}. These results are also displayed in Fig.~\ref{fig:Snowlinelocations}.

A snowline location further out than expected based on the luminosity could indicate that the source has recently undergone an accretion burst (e.g., \citealt{Lee2007,Visser2012,Jorgensen2015}). During the burst the luminosity increases, heating up the circumstellar material and shifting the snowlines outward. While the temperature adapts almost instantaneously when the protostar goes back to its quiescent mode of accretion \citep{Johnstone2013}, the chemistry needs time to react and the refreeze-out timescale can be expressed as 
\begin{equation}
\tau_{\rm{fr}} = 1 \times 10^4 \hspace{0.1cm} \mathrm{yr}  \hspace{0.1cm} \sqrt{\frac{\rm{10 \hspace{0.1cm} K}}{T_{\rm{dust}}}} \frac{10^6 \hspace{0.1cm}\rm{cm}^{-3}}{n_{\rm{H}_2}}, 
\end{equation} 
where $n_{\rm{H}_2}$ is the gas density and $T_{\rm{dust}}$ is the dust temperature \citep{Visser2012}. Because the water snowline is located at higher temperatures and hence higher densities than the CO snowline, the refreeze-out timescale for water is shorter than for CO ($\sim$100-1,000 yr versus $\sim$10,000 yr for protostellar envelope densities of $10^6-10^7$ cm$^{-3}$). Combining information on both the CO and water snowline will therefore provide a better constraint on when a burst happened. 

For some of the sources in our sample, the occurrence of a recent burst has been studied before. \citet{Frimann2017} inferred whether a burst must have occurred based on the extent of C$^{18}$O emission, and \citet{Hsieh2019} used the locations of the CO and H$_2$O snowlines as derived from N$_2$H$^+$ and HCO$^+$ emission, respectively. The luminosities required to match the C$^{18}$O and/or N$_2$H$^+$ and HCO$^+$ observations are also provided in Table~\ref{tab:Snowline}, together with the corresponding water snowline location using Eq.~\ref{eq:B07}. We will first discuss the snowline location and whether there is evidence of a recent burst for the sources in our sample (Sect.~\ref{sec:Bursts_H13CO+sources}). Next, we will estimate the average burst interval using the full sample of protostars for which sub-arcsecond resolution water observations have been presented in this work and in the literature (Sect.~\ref{sec:Bursts_Watersources}). We will discuss uncertainties on the derived burst interval in Sect.~\ref{sec:BurstDiscussion}.

\subsection{Sources with H\textsuperscript{13}CO$^+$ observations} \label{sec:Bursts_H13CO+sources}

\textit{B1-bS and B1-c.} For four of the five sources discussed in this work, that is, B1-bS, B1-c, B5-IRS1 and L1448-mm, the peak of H$^{13}$CO$^+$ emission is located at radii much larger than the expected snowline location, that is, at the radius where the snowline would be for a $\sim$100 times higher luminosity than the current luminosity. As discussed in Sects.~\ref{sec:B1-c_water} and \ref{sec:B1-bS}, H$_2^{18}$O and HDO observations for B1-c and B1-bS show that the snowline is actually much closer in. Figure~\ref{fig:Snowlinelocations} thus clearly illustrates that without any detailed source-specific analysis, the peak of the H$^{13}$CO$^+$ emission provides only an upper limit to the snowline location. For B1-c, a snowline location of $\sim$20 au is smaller than the expected location of 35 au, while a snowline at $\sim$20 au in B1-bS is larger than the expected location of 11 au. 

B1-c was also targeted in the accretion bursts studies by \citet{Frimann2017} and \citet{Hsieh2019}. The latter study concluded that B1-c has recently (within the last 1,000 yr) undergone a burst based on the location of the water snowline inferred from HCO$^+$ observations. However, based on the H$_2^{18}$O and HDO observations, and our results that the main isotopologue HCO$^+$ is not a good tracer of the snowline, we conclude that B1-c has not undergone a recent burst. \citet{Frimann2017} showed that B1-c could have undergone a burst assuming a CO freeze out temperature of 28 K, but the extent of the C$^{18}$O emission was consistent with the luminosity for a freeze-out temperature of 21 K. Combining this with the result from \citet{Hsieh2019} that N$_2$H$^+$ observations are consistent with the current luminosity, we conclude that B1-c has not undergone an accretion burst within the last 10,000 yr. 

B1-bS has not been studied before, so we can only constrain that no burst occurred during the last 1,000 yr. Here we adopt the criteria used by \citet{Jorgensen2015} and \citet{Frimann2017} that a source is classified as having recently undergone a burst if the snowline location corresponds to a luminosity $>$5 times higher than the current luminosity. A snowline at 20 au in B1-bS suggests a luminosity of only 3.2 $\times L_{\rm{current}}$, so we do not consider B1-bS a post-burst source. 

\textit{L1448-mm.} A snowline at $\sim$20 au in L1448-mm, as suggested from the similarity in H$^{13}$CO$^+$ morphology between L1448-mm and B1-c, would be $\sim$25 au closer in than expected from the luminosity. However, \citet{Maret2020} recently showed that the C$^{18}$O emission displays Keplerian rotation out to 200 au. A disk would increase the amount of dense and cold material on small scales \citep{Persson2016}, and can shield the inner envelope from the central heating \citep{Murillo2015}. A higher luminosity is therefore required to obtain a certain peak radius for HCO$^+$ when a disk is present \citep{Hsieh2019}, and Eq.~\ref{eq:B07} does not provide a good prediction of the snowline location. The models by \citet{Hsieh2019} show that the HCO$^+$ peak shifts about 30 au inward when a disk is present for a 9 $L_\odot$ star. Assuming that the HCO$^+$ peak shift in these models is representative for the snowline shift, this would be in agreement with the inferred snowline location. Adopting here the current luminosity determined by \citet{Karska2018}, the luminosity derived by \citet{Frimann2017} to match the C$^{18}$O extent provides weak evidence for an accretion burst: the luminosity needs to be increased by a factor of 5.6, and previous studies considered a threshold of a factor 5 for a significant burst \citep{Jorgensen2015,Frimann2017}. Given that the analysis by \citet{Frimann2017} did not include the presence of a disk, we tentatively conclude that L1448-mm has likely undergone a burst more than 1,000 yr but less than 10,000 yr ago. The influence of the presence of a disk on snowline locations will be investigated in more detail in Murillo et al. in prep. 

\textit{B5-IRS1.} If the relationship between the H$^{13}$CO$^+$ peak and the water snowline would be similar to that in B1-bS and B1-c, that is, H$^{13}$CO$^+$ peaking where the snowline is expected for $\sim$100 times the current luminosity, then the location of the H$^{13}$CO$^+$ peak would suggest that the snowline in B5-IRS1 is roughly at its expected location (see Fig.~\ref{fig:Snowlinelocations}). However, H$^{13}$CO$^+$ was also found to peak at a snowline radius corresponding to a 100 times increased luminosity ($\sim$200 au) in IRAS15398 \citep{Jorgensen2013}, while the extent of the spatially resolved HDO $1_{0,1}-0_{0,0}$ emission ($\sim$100 au; \citealt{Bjerkeli2016}) is consistent with a luminosity increase of a factor $\sim$25. Although there may be some uncertainty in the exact snowline location in IRAS15398 as HDO emission was also observed along the outflow cone, these observations suggest that there is not a simple uniform rule to convert the location of the H$^{13}$CO$^+$ peak into a snowline location. It is therefore hard to narrow down the snowline radius in B5-IRS1 without a more detailed study of this source. 

The luminosity needs to increase by a factor 9 to explain the C$^{18}$O extent (for a freeze-out temperature of 28 K; \citealt{Frimann2017}), which would shift the water snowline to $\sim$127 au. B5-IRS1 thus seems to have undergone a burst within the past 10,000 yr (based on C$^{18}$O), but the current observations cannot determine whether this burst happened within the past 1,000 yr. 

\textit{HH211.} At a resolution of $0.73^{\prime\prime} \times 0.58^{\prime\prime}$ the H$^{13}$CO$^+$ emission toward HH211 is centrally peaked, and only when imaged at slightly higher resolution is a tentative central depression visible with a peak at $\sim$75 au. This behavior for H$^{13}$CO$^+$ is deviating from the trend seen with luminosity (Fig.~\ref{fig:Snowlinelocations}), and sources with both lower and higher luminosity have H$^{13}$CO$^+$ emission peaking at larger radii. There is evidence for a small disk around HH211 ($\sim$10 au radius; \citealt{Segura-Cox2016,Segura-Cox2018,Lee2018}), which could be the reason for H$^{13}$CO$^+$ peaking closer in than expected. However, if the tentative depression is not real and the emission is centrally peaked, this disk would have a much stronger impact than the disk around L1448-mm. Another explanation for H$^{13}$CO$^+$ peaking on or very close to the source could be the near edge-on geometry of HH211 \citep{Gueth1999,Lee2009}. Models by \citet{Hsieh2019} showed that the there is no central gap in H$^{13}$CO$^+$ emission for highly inclined sources, although in these models the emission on one side of the continuum peak is brighter than from the other side, rather than centrally peaked. 

A third scenario could involve the destruction of water due to a high X-ray luminosity \citep{Stauber2005,Stauber2006,Notsu2021}. In particular, the chemical modeling done by \citet{Notsu2021} showed that this process would significantly increase the HCO$^+$ abundance inside the snowline and decrease the CH$_3$OH abundance. This could be consistent with centrally peaked H$^{13}$CO$^+$ emission, a lower H$_2^{18}$O column than toward B1-c and the non-detection of CH$_3$OH, but higher resolution observations including H$_2^{18}$O or HDO are required to confirm this scenario. While X-ray emission is widely observed toward T Tauri stars (e.g., \citealt{Gudel2007}), Class 0 protostars are too deeply embedded to be detected \citep{Giardino2007}. However, recently the detection of an X-ray flare was reported for the Class 0 protostar HOPS 383 \citep{Grosso2020}, suggesting that this type of emission could play a role in the chemistry of these young objects. A temporal phenomenon as flares may then explain why this effect is only observed toward HH211.  

Taken together, the H$^{13}$CO$^+$ observations toward HH211 are thus not strongly suggesting a recent ($<$ 1,000 yr) accretion burst, although this could still be possible if the water abundance is too low to affect the H$^{13}$CO$^+$ abundance. Based on the C$^{18}$O spatial extent a burst is required assuming a CO freeze out temperature of 28 K, so HH211 may have undergone a burst 1,000--10,000 yr ago.  

\subsection{Sources with H$_2$\textsuperscript{18}O or HDO observations} \label{sec:Bursts_Watersources}

Because H$^{13}$CO$^+$ observations are not stringent enough to constrain the occurence of an accretion burst, and to obtain a sample as large as possible, we compile an overview of snowline locations for all protostars with reported sub-arcsecond resolution water observations. A description of how the snowline estimates are obtained is given in Appendix~\ref{app:literaturewater} and the results are presented in Fig.~\ref{fig:Snowlinelocations} and listed in Table~\ref{tab:Snowline}.  

Comparing the derived snowline locations with the expected location based on luminosity (Eq.~\ref{eq:B07}) in Fig.~\ref{fig:Snowlinelocations} shows that for only one source (IRAS15398) the snowline is at a substantially larger radius than expected (that is, at a radius requiring $>5 \times L_{\rm{current}}$). Given its high current luminosity, IRAS2A is most likely currently in a burst phase (see also the discussion in \citealt{Hsieh2019}). Excluding IRAS2A, there is then one source out of nine that shows signs of a recent accretion burst ($<$ 1,000 yr) if we only consider sources with water observations. As discussed in Sect.~\ref{sec:Bursts_H13CO+sources}, HH211 and L1448-mm have likely not undergone a burst in the last 1,000 yr, which would mean that one out of 11 sources is showing signs of burst activity. The time interval between bursts can be estimated from the ratio between the re-freezeout timescale and the fraction of post-burst sources. These results then suggest a burst interval of 9,000--11,000 yr. 

The burst results for IRAS4A-NW are a little uncertain, because reported luminosities are for the IRAS4A binary, while the water emission peaks at the northern source IRAS4A-NW (also referred to as 4A2). If the luminosity of IRAS4A-NW is less than half of the total luminosity, it would qualify as a post-burst source. The resulting burst interval would then be 4,500--5,500 yr. 

Previous studies using C$^{18}$O emission showed that IRAS4A and IRAS4B may have undergone a burst within the last 10,000 yr \citep{Jorgensen2015,Frimann2017}, so in combination with the water observations the burst occurrence can be constrained to between 1,000-10,000 yr ago. For B335, the measured size of the C$^{18}$O emitting region did not suggest a recent burst \citep{Jorgensen2015}, and a burst is even less likely with the increased luminosity as result of a larger distance (165 versus 100 pc; \citealt{Watson2020}). Excluding the currently in burst source IRAS2A, out of the seven sources in our sample that have constraints on accretion bursts from C$^{18}$O observations, five show signs of a burst within the past 10,000 yr. This corresponds to an estimated burst interval of 14,000 yr. The burst interval derived from water observations is thus very similar to the interval derived from C$^{18}$O observations, but both numbers have a large uncertainty due to the small sample size. 

\subsection{Discussion of the burst interval} \label{sec:BurstDiscussion}

In addition to the small sample size there are other factors that contribute to the uncertainty of the estimated burst interval. One aspect is the potential presence of a disk. As discussed in Sect.~\ref{sec:Bursts_H13CO+sources} for L1448-mm, a disk would result in a snowline location closer to the star than predicted based on the luminosity using Eq.~\ref{eq:B07}. This is clearly the case for L1448-mm and L483 as their snowline location is consistent with a luminosity $< 1/5$ of the current luminosity, and suggested for HH211 by the centrally peaked H$^{13}$CO$^+$ emission. B335 is a borderline case with its smallest snowline estimate consistent with a luminosity exactly 5 times smaller than the current luminosity. If a disk is present in these sources, more detailed studies are required to determine whether the snowline location is where it is expected to be or whether a recent burst occurred.

However, we can make a first assessment using the embedded disks models presented by \citet{Harsono2015}. That study calculated the temperature structure of the disk and envelope around accreting protostars within the framework of two-dimensional physical and radiative transfer models, and used that to determine snowline locations. In these models, the water snowline location is dependent on the accretion rate for accretion rates $> 10^{-6} M_\odot$ yr$^{-1}$. Disk radii between 50 and 200 au are modelled, and the snowline lies always in the disk. The snowline ranges between $\sim$10--35 au for a 5 $L_\odot$ star, and between $\sim$15--30 au for a 15 $L_\odot$ star. The luminosities of L1448-mm and L483 are $\sim$10 $L_\odot$, and their snowline estimates of $\sim$20 au and $\sim$14--22 au, respectively, fall within the range of model predictions. A snowline radius $<$ 20 au for HH211 (3.6 $L_\odot$) would also be consistent with a disk in these models. Higher accretion rates shift the snowline further outward for lower luminosities, and for a 1 $L_\odot$ star, the models predict a snowline location between 5 and 55 au. The results for B335 (2 $L_\odot$ and a snowline at 10--14 au) could thus also be consistent with the presence of a disk. If these four sources actually have a disk then these models do not point to a recent burst. This assessment does not change our burst estimate as we are already assuming that these sources have not recently undergone a burst. 

While sources with a snowline location closer in than expected thus do not suggest a recent burst, we cannot rule out that sources that have a snowline location consistent with their current luminosity are in fact sources with a disk that have recently undergone an accretion burst. In order to properly classify a source as having recently undergone a burst or not, high-resolution (spatially and spectrally) molecular line observations are needed to establish whether a disk is present or not. 

Large uncertainties in the locations of the snowline as is the case for sources with only H$^{13}$CO$^+$ observations will also contribute to dispersions in the estimated burst interval, especially with this small sample size. In the current sample, B5-IRS1 is the only source with a large uncertainty in snowline radius. If B5-IRS1 would be added as a quiescent source, the burst interval would increase slightly from 9,000-11,000 yr to 10,000-12,000 yr, and adding B5-IRS1 as a post-burst source would decrease the burst interval by a factor of almost two to 5,000-6,000 yr.

Another caveat in the analysis of accretion bursts by comparing snowline location to source luminosity is that the luminosity for edge-on sources may be substantially greater than the observed value. However, a higher luminosity would make a burst less likely for an observed snowline location. The only source in our sample that could be affected is IRAS15398 as the current luminosity suggests that a burst has occurred in the past 1,000 yr. An inclination angle of 20$^\circ$ derived from the outflow indicates that this source is viewed nearly edge-on \citep{Oya2014}. However, in order for this source to be classified as not having undergone a recent burst, its luminosity would need to be larger than $\sim$7$L_\odot$; its luminosity is currently determined to be 1.8 $L_\odot$ \citep{Jorgensen2013}. 

Having an additional indicator of a recent burst would help mitigate the uncertainties discussed above. Chemical modeling has shown that the ice evaporation induced by an accretion burst could trigger gas-phase formation of complex organic molecules (COMs; \citealt{Taquet2016}). Strong COM emission may thus be an indicator of a recent burst. Recently, \citet{Yang2021} detected COM emission toward 58\% of sources in a chemical survey of 50 protostars in Perseus. They found no relationship between COM emission and bolometric luminosity, but the study did not address the effect of accretion history. If all sources with COMs are post-burst and the bursts happened less than 1,000 year ago, this would suggest a burst interval of 1,700 yr. From the sample presented in this work, B1-c shows the strongest COM emission followed by L1448-mm, while no COMs were detected toward HH211 and B5-IRS1. Since there is no clear evidence for a burst in the last 10,000 yr in B1-c, while C$^{18}$O observations point to a burst less than 10,000 yr ago for the other three sources, the use of COM emission as burst indicator is not evident from this sample. In addition, COM emission toward the protostar HH212 has been suggested to be related to an accretion shock at the disk-envelope interface \citep{Lee2017,Codella2018}. More studies are thus required to determine if and how COM emission relates to accretion bursts. 

The burst interval estimate from the water snowline, on the order of $\sim$10,000 yr, falls in between previous estimates. \citet{Jorgensen2015} found a burst interval of 20,000--50,000 yr based on C$^{18}$O observations of 16 protostars, similar to the results from \citet{Frimann2017} for a sample of 19 sources. On the other hand, \citet{Hsieh2019} derive a burst interval of $\sim$2,400 and 8,000 yr for Class 0 and Class I protostars, respectively. Given the small number of sources with water observations it is hard to rule out a burst interval longer than $\sim$10,000 yr. A burst interval of only 2,400 yr seems unlikely if only one out of 11 sources are found to be in the post-burst stage. Assuming a binomial distribution this chance is only $\sim$2\%. If both IRAS4A-NW and B5-IRS1 have undergone a burst in the last 1,000 yr a burst interval of 2,400 yr becomes slightly more likely ($\sim$17\% chance of finding three out of 11 sources in post-burst stage). 

We adopted a timescale of 1,000 yr for the refreeze out of water, as done by \citet{Hsieh2019}. This corresponds to a density of $\sim 10^{6}$ cm$^{-3}$. For densities an order of magnitude higher, this timescale decreases to $\sim$100 yr. Inner envelope densities $> 10^{7}$ cm$^{-3}$ are not unlikely (e.g., \citealt{Kristensen2012}), especially when a disk is present. In case of a disk, densities of $\sim 10^{6}$ cm$^{-3}$ may still be appropriate if the water emission arises predominantly from surface layers. Nonetheless, shorter refreeze-out timescales would result in a shorter burst-interval estimate. For a freeze-out timescale of 100 yr our burst interval estimate would lower by a factor 10 to 900--1100 yr, and the results from \citet{Hsieh2019} would lower to 240 yr for Class 0 and 800 yr for Class~I.

In order to better constrain the burst frequency, we thus need high-resolution water observations of a large number of sources that provide a good representation of the protostellar population. The current sample of protostars with water observations consists mainly of the more luminous objects and is dominated by targets in Perseus. In addition, a detailed characterization of the inner region is required to determine whether a disk is present or not and to constrain the refreeze-out timescale.


\section{Conclusions} \label{sec:Conclusions}

We have presented a suite of molecular line observations (H$_2^{18}$O, HDO, HCO$^+$, H$^{13}$CO$^+$, and HC$^{18}$O$^+$) at $\sim$0.2$^{\prime\prime}-$0.7$^{\prime\prime}$ (60--210 au) resolution to study the water snowline and the occurrence of accretion bursts in protostellar envelopes. Our main conclusions are the following:

\begin{itemize}
\item The compact H$_2^{18}$O and HDO emission surrounded by a ring of H$^{13}$CO$^+$ $J=2-1$ and HC$^{18}$O$^+$ $J=3-2$ toward B1-c provides a textbook example of a chemical snowline tracer. Deconvolving the water emission results in a snowline estimate of 19 $\pm$ 6 au, well within the peak of the H$^{13}$CO$^+$ emission at 300 au. Similar results are found for HDO and H$^{13}$CO$^+$ $J=3-2$ toward B1-bS. 

\item The main isotopologue HCO$^+$ is not suited to trace the water snowline in protostellar envelopes because the emission is optically thick. The best H$^{13}$CO$^+$ line is the $J=2-1$ transition, because the $J=3-2$ and $J=4-3$ transitions can be blended with lines from complex organics and the $J=1-0$ transition will be dominated by emission from colder material in the outer envelope. However, the H$^{13}$CO$^+$ emission peak provides, at best, an upper limit to the water snowline location. This corroborates earlier results that in order to derive a snowline location from H$^{13}$CO$^+$ emission several factors have to be taken into account, such as the fact that the H$^{13}$CO$^+$ column peaks slightly outside of the snowline, the inclination of the source, the presence of a disk, absorption by larger-scale material, the beam size of the observations, and possibly X-ray flares. The inner edge of HC$^{18}$O$^+$ emission may provide a stronger constraint on the snowline location. 

\item There is no evidence of an accretion burst during the last $\sim$1,000 yr in B1-bS, B1-c, HH211 and L1448-mm, while this cannot be ruled out for B5-IRS1. The anticipated relation between the water snowline location and the source luminosity is clearly present in the dataset compiled from all existing sub-arcsecond resolution observations of water and H$^{13}$CO$^+$ towards protostars. One out of 11 sources is showing signs of burst activity in the past 1,000 yr and we derive an average burst interval on the order of $\sim$10,000 yr. However, water observations for a larger source sample are required for a better constraint. 

\item The HDO/H$_2$O ratio in B1-c is found to be (7.6 $\pm$ 0.9) $\times$ 10$^{-4}$, very similar to the ratios derived toward four other protostars in Perseus and Ophiuchus.

\end{itemize}

Given the extended analysis required to derive a snowline location from H$^{13}$CO$^+$ or HC$^{18}$O$^+$, their value lies in the application in sources where water cannot be readily detected, such as circumstellar disks. The most straightforward way to locate the water snowline in protostellar envelopes is through direct observations of H$_2^{18}$O or HDO.


\acknowledgments 

The authors would like to thank the referee for constructive comments and Neal Evans for commenting on the manuscript. This paper makes use of the following ALMA data: ADS/JAO.ALMA\#2016.1.00505.S, \\ ADS/JAO.ALMA\#2017.1.00693.S, \\ ADS/JAO.ALMA\#2017.1.01174.S, \\ ADS/JAO.ALMA\#2017.1.01371.S, \\ ADS/JAO.ALMA\#2017.1.01693.S, \\ ADS/JAO.ALMA\#2019.1.00171.S and \\ ADS/JAO.ALMA\#2019.1.00720.S. ALMA is a partnership of ESO (representing its member states), NSF (USA) and NINS (Japan), together with NRC (Canada), MOST and ASIAA (Taiwan), and KASI (Republic of Korea), in cooperation with the Republic of Chile. The Joint ALMA Observatory is operated by ESO, AUI/NRAO and NAOJ. Astrochemistry in Leiden is supported by the Netherlands Research School for Astronomy (NOVA). M.L.R.H acknowledges support from a Huygens fellowship from Leiden University and the Michigan Society of Fellows. E.A.B. acknowledges support from NSF AAG Grant \#1907653. J.K.J acknowledges support from the Independent Research Fund Denmark (grant number DFF0135-00123B) J.J.T acknowledges support from  NSF AST-1814762. The National Radio Astronomy Observatory is a facility of the National Science Foundation operated under cooperative agreement by Associated Universities, Inc. D.H. acknowledges support from the EACOA Fellowship from the East Asian Core Observatories Association. M.G. acknowledges support from the Dutch Research Council (NWO) with project number NWO TOP-1 614.001.751.


\bibliography{References}{}
\bibliographystyle{aasjournal}




\restartappendixnumbering

\begin{appendix}

\section{ALMA observing log}

Table~\ref{tab:Observations} presents details of the different ALMA observations used in this work. 

\vspace{0.5cm}

\begin{deluxetable*}{llcccccc}
\tablecaption{ALMA observing log. \label{tab:Observations}}
\tablewidth{0pt}
\addtolength{\tabcolsep}{-1pt} 
\tabletypesize{\scriptsize}
\tablehead{
\colhead{ALMA program} & Date & Bandpass calibrator & Flux calibrator & Phase calibrator  & Max. baseline (km) & $N_{\rm{antenna}}$ & ALMA band
} 
\startdata 
2016.1.00505.S		& 2016 Oct 8			& J0237+2848					& J0238+1636		& J0336+3218				& 3.1							& 43			& 6 \\
							& 2016 Oct 13			& J0237+2848					& J0238+1636		& J0336+3218				& 3.1 						& 44 		& 6 \\
							& 2016 Oct 14			& J0237+2848					& J0238+1636		& J0336+3218				& 2.5 						& 46			& 6 \\
2017.1.00693.S		& 2018 Jan 15			& J0904--5735					& J0904--5735		& J1147--6753				& 2.4						& 46			& 6 \\ 
							& 2018 Mar 11			& J1751+0939					& J1751+0939		& J1743--0350				& 1.2							& 42			& 6 \\
							& 2018 Mar 20			& J2025+3343					& J2025+3343		& J1955+1358				& 0.74						& 44			& 6 \\ 
							& 2018 Aug 27			& J2000--1748					& J2000--1748		& J1938+0448				& 0.78						& 45			& 5 \\
							& 2018 Aug 27			& J1751+0939					& J1751+0939		& J1743--0350				& 0.78						& 44			& 5 \\
							& 2018 Sep 4			& J1107--4449					& J1107--4449		& J1147--6753				& 0.78						& 43			& 5 \\
2017.1.01174.S		& 2018 Sep 7			& J0237+2848					& J0237+2848		& J0336+3218				& 0.78						& 47			& 6 \\
2017.1.01371.S		& 2018 Sep 16			& J0237+2848					& J0237+2848		& J0336+3218				& 1.3						 	& 45			& 5 	\\
							& 2018 Sep 25			& J0237+2848					& J0237+2848		& J0336+3218				& 1.4							& 45			& 5 \\
2017.1.01693.S 	& 2018 Sep 15			& J0237+2848					& J0237+2848		& J0336+3218				& 1.3							& 44			& 6 \\
							& 2018 Sep 16			& J0237+2848					& J0237+2848		& J0336+3218				& 1.3							& 45			& 6 \\
							& 2018 Sep 17			& J0237+2848					& J0237+2848		& J0336+3218				& 1.2							& 45			& 6 \\
							& 2018 Sep 20			& J0237+2848					& J0237+2848		& J0336+3218				& 1.4							& 47			& 6 \\
							& 2018 Sep 21			& J0237+2848					& J0237+2848		& J0336+3218				& 1.4							& 43			& 6 \\
							& 2018 Sep 22			& J0237+2848					& J0237+2848		& J0336+3218				& 1.4							& 44			& 6 \\
2019.1.00171.S		& 2019 Oct 22			& J0510+1800					& J0510+1800		& J0336+3218		 		& 0.78						& 47			& 5	\\
							& 2019 Oct 23			& J0237+2848					& J0237+2848		& J0336+3218				& 0.78						& 47			& 5	\\
2019.1.00720.S		& 2019 Oct 8			& J1924--2914					& J1924--2914		& J1938+0448				& 0.78 						& 42 		& 7 \\
							& 2019 Oct 29			& J1924--2914					& J1924--2914		& J1743--0350				& 0.70 						& 45 		& 7 \\ 
\enddata
\end{deluxetable*}

\section{Continuum images} \label{app:Continuum}

Figure~\ref{fig:Continuum} presents continuum images of the protostellar envelopes in our sample. These images are corrected for the primary beam to make sure that the fluxes are correct. This is particularly important for B1-bS which is near the edge of the primary beam in this dataset. 

\begin{figure*}
\centering
\includegraphics[width=\textwidth,trim={0cm 16.2cm 0cm 2.0cm},clip]{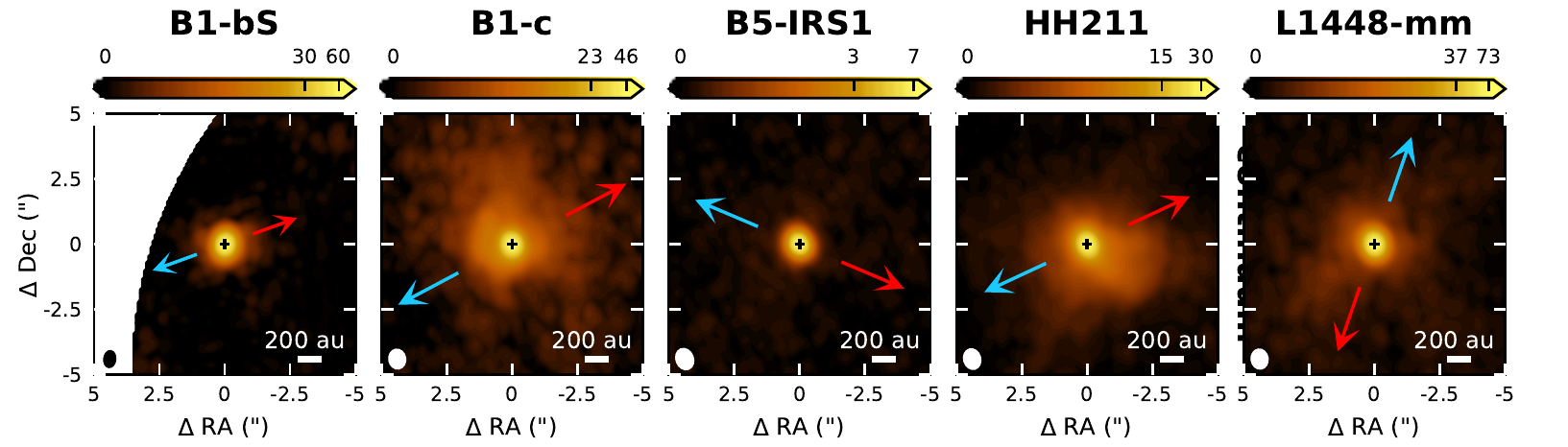}
\caption{Continuum images at 1.2 mm for B1-bS and at 1.7 mm for the other protostars in our sample. The color scale is in mJy beam$^{-1}$. The outflow directions are indicated by blue and red arrows, the continuum peak is marked by a cross and the beam is shown in the lower left corner of each panel.}
\label{fig:Continuum}
\end{figure*}

\section{HDO upper limit for B1-bN} \label{app:B1-bN}

B1-bN was only targeted in the HDO $3_{1,2}-2_{2,1}$ observations, and the line was not detected. We can thus not discuss the snowline location in this source, but for completeness we determine the upper limit for the HDO column density using Eqs.~\ref{eq1} and \ref{eq2}. The rms in the spectrum extracted in the central beam amounts to 0.96 mJy. Assuming a line width of 3.5 km s$^{-1}$ as observed for B1-c and an excitation temperature of 124~K, this results in an upper limit for the HDO column density of $1.1 \times 10^{14}$ cm$^{-2}$. A more narrow line width of 1 km s$^{-1}$ as observed toward B1-bS would lower the column by a factor of $\sim$2. Based on the luminosity of 0.26 $L_\odot$, the snowline is expected at $\sim$8 au. Adopting the area inside this snowline radius as source size results in an upper limit of $3.4 \times 10^{15}$ cm$^{-2}$, $\sim$2 times lower than observed toward B1-bS and $\sim$25 times lower than toward B1-c.

\section{Additional H$^{13}$CO$^+$ spectra and images} \label{app:Lineprofiles}

Figure~\ref{fig:LineprofilesH13COp-21} presents spectra of the H$^{13}$CO$^+$ $J=2-1$ transition toward B1-c, B5-IRS1, HH211 and L1448-mm. Channels with redshifted emission toward B1-c and blueshifted emission toward L1448-mm are excluded when creating moment zero maps and radial profiles. The moment zero maps of H$^{13}$CO$^+$ $J=2-1$ as presented in Fig.~\ref{fig:H13COp-21} are shown on a larger scale in Fig.~\ref{fig:largescale}. This figure also includes the images for HCO$^+$ and its isotopologues toward B1-c. Finally, Fig.~\ref{fig:Spectra_B1-bS} shows the H$^{13}$CO$^+$ $J=3-2$ spectrum toward B1-bS. Channels with absorption are excluded when creating the moment zero map in Fig.~\ref{fig:B1-bS}.

\begin{figure*}
\centering
\includegraphics[width=\textwidth,trim={0cm 14.4cm 0cm 1.7cm},clip]{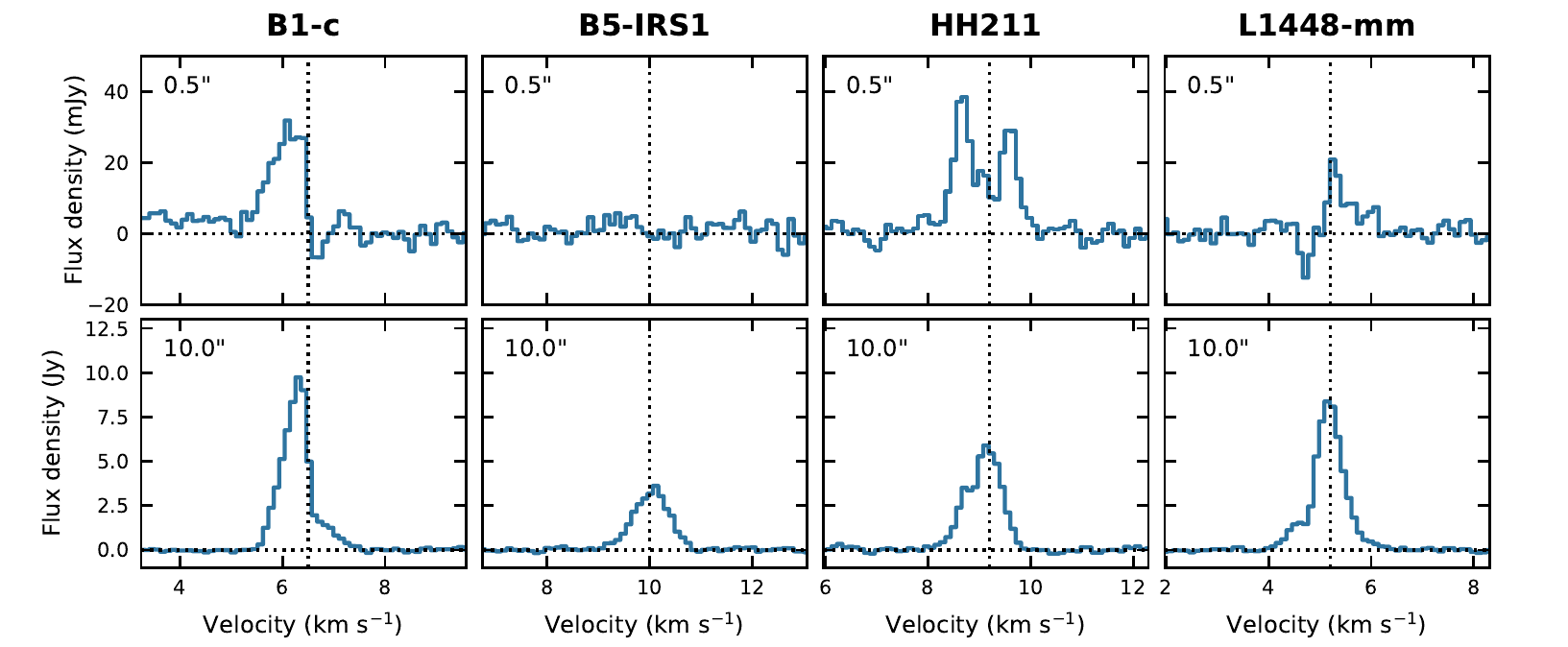}
\caption{Spectra of the H$^{13}$CO$^+$ $J=2-1$ transition toward B1-c, B5-IRS1, HH211 and L1448-mm. The spectra in the top panels are extracted within a circular 0.5$^{\prime\prime}$ diameter aperture ($\sim$one beam) centered on source. The spectra in the bottom panels are extracted in a 10$^{\prime\prime}$ aperture. The horizontal dotted line marks the zero flux level, and the vertical dotted line represents the source velocity. }
\label{fig:LineprofilesH13COp-21}
\end{figure*}

\begin{figure*}
	\centering
	\subfloat{\includegraphics[trim={0cm 15.8cm 0cm 1.4cm},clip]{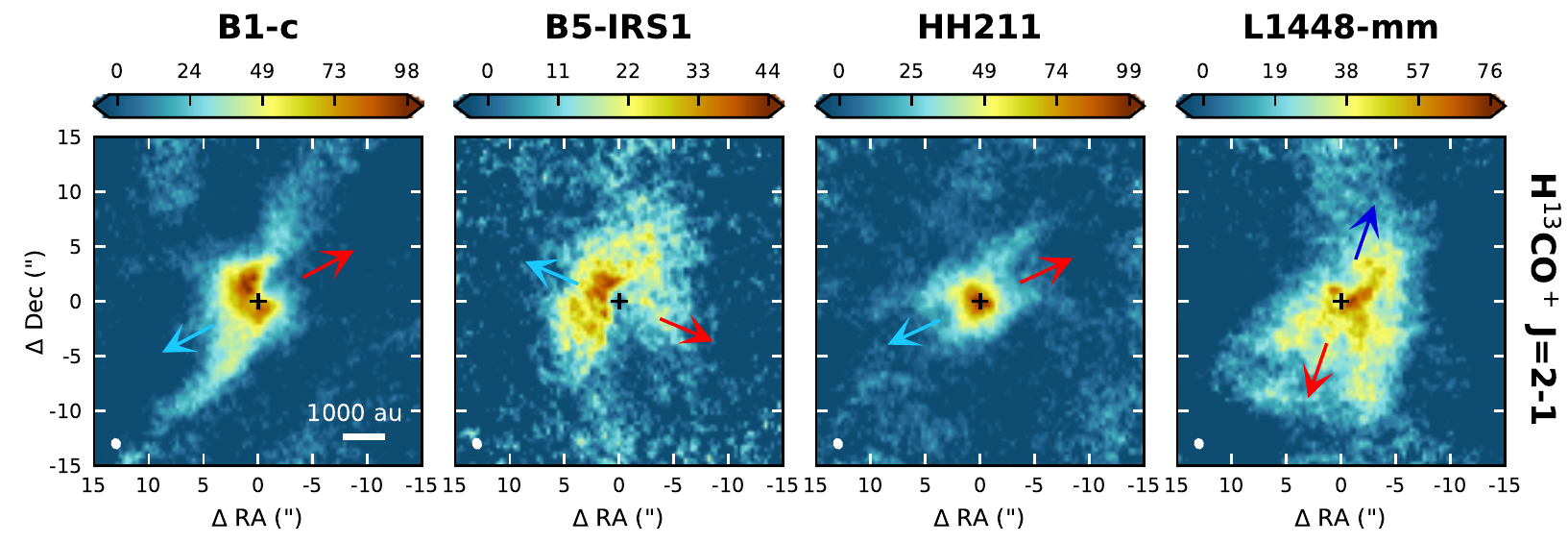}}
	\hfill
	\subfloat{\includegraphics[trim={0cm 15.8cm 0cm 1.4cm},clip]{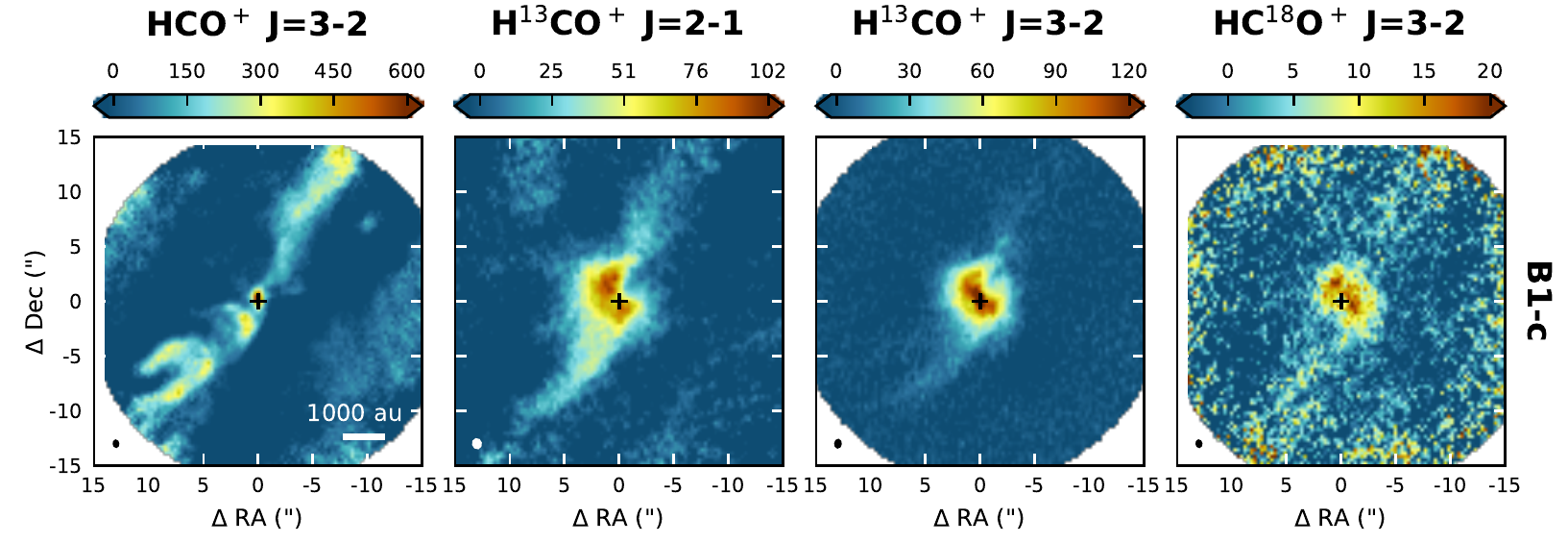}}
	\caption{Top panels: Integrated intensity maps for H$^{13}$CO$^+$ $J=2-1$ as shown in Fig.~\ref{fig:H13COp-21} but on a larger scale. Bottom panels: Integrated intensity maps for the HCO$^+$ $J=3-2$, H$^{13}$CO$^+$ $J=2-1$, H$^{13}$CO$^+$ $J=3-2$ and HC$^{18}$O$^+$ $J=3-2$ transitions toward B1-c as shown in Fig.~\ref{fig:B1-bS} but on a larger scale. The color scale for H$^{13}$CO$^+$ $J=3-2$ is adapted to better show the more extended emission. The black cross marks the continuum peak and the color scale is in mJy beam$^{-1}$ km s$^{-1}$. The beam size is depicted in the lower left corners and the outflow directions are marked by blue and red arrows in the top panels. }
\label{fig:largescale}
\end{figure*}

\begin{figure}
\centering
\includegraphics[trim={0.5cm 15.6cm 8cm 1.5cm},clip]{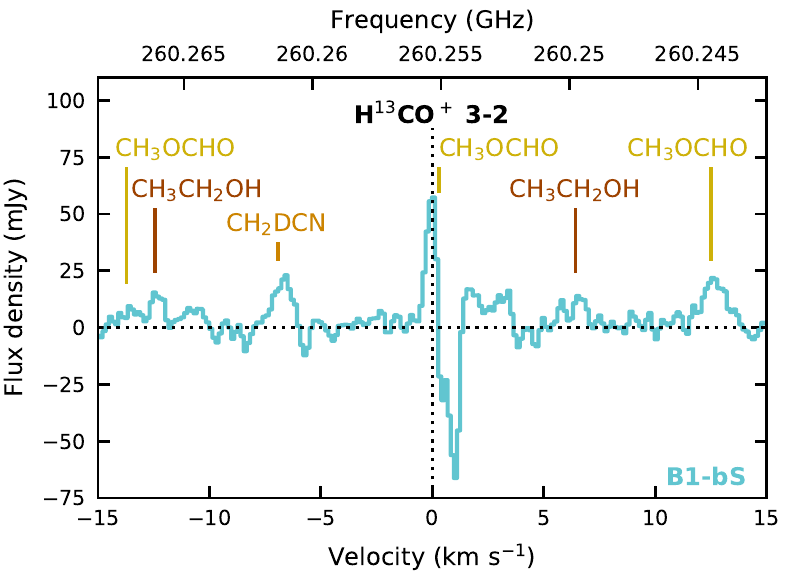}
\caption{Spectra extracted within a circular 0.5$^{\prime\prime}$ diameter aperture ($\sim$one beam) toward the continuum peak position of B1-bS centered around the H$^{13}$CO$^+$ $J=3-2$ transition.}
\label{fig:Spectra_B1-bS}
\end{figure}

\section{Methanol emission toward B1-c and L1448-mm} \label{app:CH3OH}

Figure~\ref{fig:CH3OH} presents moment zero maps of the CH$_3$OH $15_{1,15}-14_{ 2,12}$ transition at 187.5429 GHz ($E_{\rm{up}}$ = 290 K) toward B1-c and L1448-mm overlaid on the H$^{13}$CO$^+$ $J=2-1$ moment zero maps. 

\begin{figure}
\centering
\includegraphics[trim={0cm 15.8cm 8cm 1.5cm},clip]{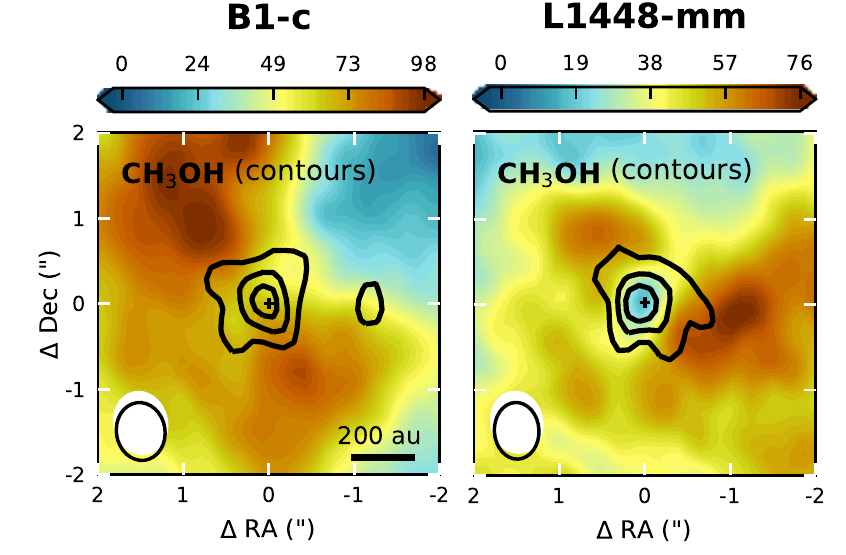}
\caption{Integrated intensity maps for the CH$_3$OH $15_{1,15}-14_{ 2,12}$ transition at 187.5429 GHz (contours) toward B1-c and L1448-mm overlaid on the integrated intensity maps for H$^{13}$CO$^+$ $J=2-1$ (color scale). Contours start at 3$\sigma$ and are in steps of 3$\sigma$, where $\sigma$ $\sim$ 20 mJy beam$^{-1}$ km s$^{-1}$. The beam is depicted in the lower left corner and the position of the continuum peak is indicated by a black cross. }
\label{fig:CH3OH}
\end{figure}

\section{Snowline location in high- versus low-mass protostellar envelopes} \label{app:high-low}

\citet{Bisschop2007} used the 1D dust radiative transfer code DUSTY \citep{Ivezic1997} to derive the radius at which the temperature reaches 100 K in high-mass protostellar envelopes (Eq.~\ref{eq:B07}). In these models, the luminosity ($10^4-10^6 L_\odot$) is provided by a single 30,000 K blackbody and the envelope has a power-law density profile, $n \propto r^{-1.5}$, typical of a free-falling core. The mass of the envelope is adjusted to match observed SCUBA 850 $\mu$m fluxes. To test if this relation holds in the low-mass regime, we ran a similar set of 1D radiative transfer models for luminosities in the range $\sim$0.5--50 $L_\odot$ using TRANSPHERE \citep{Dullemond2002}\footnote{The TRANSPHERE code is hosted online at https://www.ita.uni-heidelberg.de/~dullemond/software/transphere/index.shtml}. We varied the density power-law index between 1.5 and 2.0, that is, from free fall to a singular isothermal sphere, and the envelope mass between 0.5 and 5 $M_\odot$. The results are presented in Fig.~\ref{fig:high-low}. 

The average relation for low-mass envelopes is remarkably similar to the relation derived for high-mass sources. The slope is slightly shallower than predicted from the high-mass models, but the 100 K radius differs by only a few au for a given luminosity. In fact, the intrinsic uncertainty on the snowline location due to uncertainties in the envelope profiles (the blue shaded area in Fig.~\ref{fig:high-low}) is larger (20-30\%) than the difference between the average low-mass case and the high-mass relation. The spread in snowline radius for the low-mass sources is mainly due to the exact location where the envelope becomes optically thick to its own radiation. For more massive envelopes, the radiation is trapped at smaller scales, pushing the 100 K radius slightly further out than in a lower mass envelope. Counter-intuitively, the slope of the relation derived for high-mass protostars, which have more massive envelopes, is closer to the slope of the optically thin case than the relation derived for low-mass protostars. This is likely because their larger luminosities ($10^4-10^6 L_\odot$) push the 100 K snowline out to large enough radii that the envelope becomes optically thin to its own radiation again. 

\begin{figure}
\centering
\includegraphics[width=\linewidth,trim={0cm 0cm 11cm 0cm},clip]{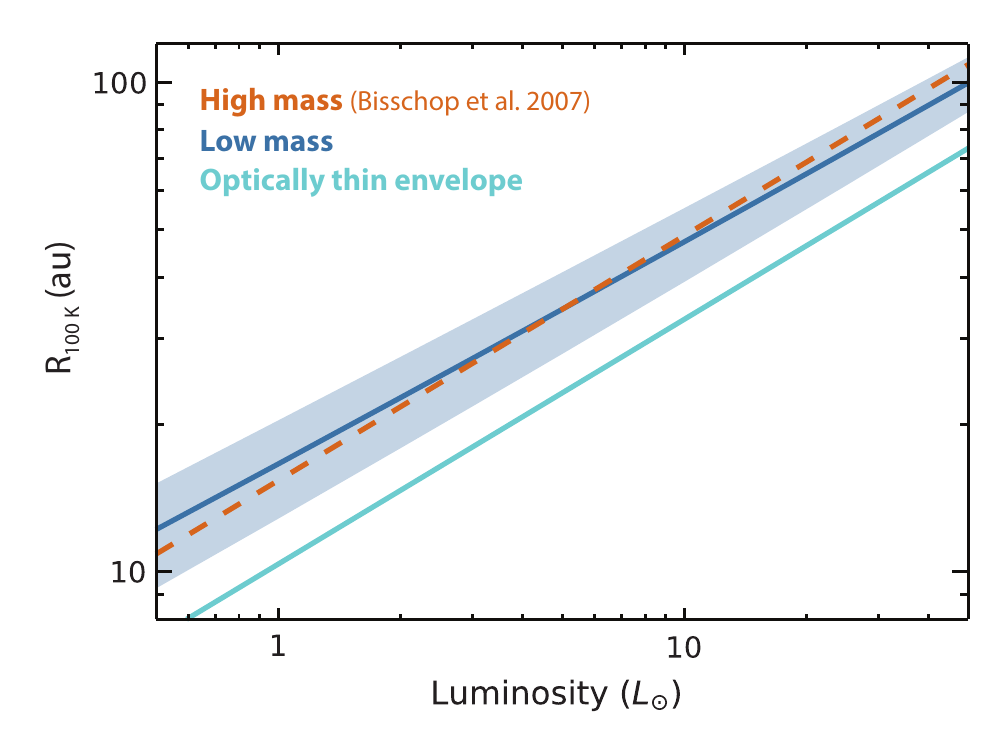}
\caption{Radius at which the temperature reaches 100 K, that is, the water snowline, in 1D radiative transfer models of low-mass protostars (solid dark blue line) compared to the relation derived by \citet{Bisschop2007} for high-mass protostars with luminosities of $10^4-10^6 L_\odot$ (Eq.~\ref{eq:B07}; dashed orange line). The shaded blue area represents the spread in 100 K radius for low-mass protostars due to varying the envelope mass between 0.5 and 5 $M_\odot$ and the density power-law index between 1.5 and 2.0. The light blue line presents the case of a completely optically thin envelope \citep{Chandler2000}.} 
\label{fig:high-low}
\end{figure}

\section{Snowline locations from sub-arcsecond water observations in the literature} \label{app:literaturewater}

\citet{Persson2013} present H$_2^{18}$O $3_{1,3}-2_{2,0}$ and HDO $3_{1,2}-2_{2,1}$ observations toward IRAS16293A and fit a circular Gaussian in the \textit{u,v}-plane. We take 0.5 $\times$ FWHM of the best fit Gaussian as an estimate of the snowline, which corresponds to 102 $\pm$ 42 au and 66 $\pm$ 24 au for H$_2^{18}$O and HDO, respectively. A similar analysis has been done for NGC1333 IRAS2A, NGC1333 IRAS4A-NW and NGC1333 IRAS4B \citep{Persson2012,Persson2014}. The H$_2^{18}$O emission toward IRAS2A has a contribution from the southern outflow lobe, and a radius of 125 $\pm$ 7.5 au based on the Gaussian fit may overestimate the snowline radius. The HDO $3_{1,2}-2_{2,1}$ and $2_{1,1}-2_{1,2}$ lines suggest a snowline around 60 $\pm$ 15 au and 75 $\pm$ 15 au, respectively. For IRAS4A-NW and IRAS4B, it is the HDO $3_{1,2}-2_{2,1}$ emission that shows outflow contributions, so we use the H$_2^{18}$O extent to get a snowline radius of 92 $\pm$ 18 and 30 $\pm$ 15 au for IRAS4A-NW and IRAS4B, respectively. 

Observations of H$_2^{18}$O $3_{1,3}-2_{2,0}$, HDO $3_{1,2}-2_{2,1}$ and HDO $2_{1,1}-2_{1,2}$ have been reported toward the isolated protostars B335, L483 and BHR71-IRS1 \citep{Jensen2019}, as well as observations of D$_2$O $1_{1,0}-1_{0,1}$ toward B335 and L483 \citep{Jensen2021}. These studies do not report source sizes, so we fit Gaussians in the image plane using the CASA \textit{imfit} task and use 0.5 $\times$ FWHM of the minor axis as a snowline estimate, as done for B1-c and B1-bS. For B335, all four lines give very similar results, and suggest a snowline radius of $\sim$10--14 au. The H$_2^{18}$O emission toward L483 is unresolved, but the HDO and D$_2$O lines suggest a snowline radius between $\sim$14--22 au. The HDO line profiles toward BHR71-IRS1 show slight deviations from a Gaussian profile which could be related to weak emission from other components than the inner envelope, and the H$_2^{18}$O line is partly blended. This may explain why the difference in estimated snowline location from both isotopologues is larger (44$\pm$13 au versus 24$\pm$3 au). All snowline estimates are listed in Table~\ref{tab:LiteratureWater}. 

A comparison using the HDO $2_{1,1}-2_{1,2}$ observations toward IRAS2A shows that a snowline estimate based on the minor axis of an elliptical Gaussian in the image plane is comparable to an estimate based on a circular Gaussian fit in the (\textit{u,v,})-plane. The former method returns a Gaussian with major and minor axes of 137 $\pm$ 33 au and 98 $\pm$ 30 au, respectively. The latter method gives a FWHM of 120 $\pm$ 30 au. In addition to these $1.4^{\prime\prime} \times 0.9^{\prime\prime}$ NOEMA observations, HDO $2_{1,1}-2_{1,2}$ has been observed toward IRAS2A with ALMA at $0.074^{\prime\prime} \times 0.035^{\prime\prime}$ resolution (archival ALMA data, project code 2018.1.00427.S). The Gaussian fit to the marginally resolved NOEMA observations results in a deconvolved Gaussian ($0.46^{\prime\prime} (\pm 0.11^{\prime\prime}) \times 0.33^{\prime\prime} (\pm 0.09^{\prime\prime})$) that agrees within the errorbars with the better constrained result from the highly resolved ALMA observations ($0.40^{\prime\prime} (\pm 0.07^{\prime\prime}) \times 0.27^{\prime\prime} (\pm 0.05^{\prime\prime})$). As long as the emission is marginally resolved, we are thus able to derive an approximate snowline location.

\begin{deluxetable*}{lcccccl}
\tablecaption{Overview of H$_2$O snowline locations based on observations presented in the literature. \label{tab:LiteratureWater}}
\tablewidth{0pt}
\tabletypesize{\scriptsize}
\tablehead{
\colhead{Source} 	& \multicolumn{5}{c}{Snowline radius (au)} & \colhead{Reference} \\\hline
\colhead{}	 \vspace{-0.3cm}	& \colhead{H$_2^{18}$O} & 
\colhead{HDO} 	& \colhead{HDO} & \colhead{HDO} & 
\colhead{D$_2$O} & \colhead{} \\
\colhead{} \vspace{-0.5cm} & \colhead{$3_{1,3}-2_{2,0}$} & 
\colhead{$3_{1,2}-2_{2,1}$} & \colhead{$2_{1,1}-2_{1,2}$} &  \colhead{$1_{0,1}-0_{0,0}$} &
\colhead{$1_{1,0}-1_{0,1}$} & \colhead{}\\ 
} 
\startdata 
B335 			& 9.5 $\pm$ 20		& 13 $\pm$ 4.8		& 14 $\pm$ 4.7 	&  -					& 10 $\pm$ 5.3 	& \citet{Jensen2019,Jensen2021} \\
BHR71-IRS1	& 44 $\pm$ 13		& 24 $\pm$ 3.1		& 24 $\pm$ 2.7 	&  -					& -						& \citet{Jensen2019,Jensen2021} \\
L483				& unresolved 		& 17 $\pm$ 7.8		& 22 $\pm$ 4.7		&  - 					& 14 $\pm$ 12		& \citet{Jensen2019,Jensen2021} \\
IRAS15398	& -						& - 						& -						& $\sim$100		& -						& \citet{Bjerkeli2016} \\
IRAS16293A	& 102 $\pm$ 42 	& 66 $\pm$ 24 	& - 						& - 					& - 						& \citet{Persson2013} \\
IRAS2A			& 125 $\pm$ 7.5	& 60 $\pm$ 15 		& 75 $\pm$ 15 		& 	-					& - 						& \citet{Persson2012,Persson2014} \\
IRAS4A-NW	& 92 $\pm$ 18 		& 240 $\pm$ 30	& - 						& -					& - 						& \citet{Persson2014} \\
IRAS4B 		& 30 $\pm$ 15		& 75 $\pm$ 15		& - 						& -					& - 						& \citet{Jorgensen2010,Persson2014} 
\enddata
\tablecomments{For B335, BHR71-IRS1 and L483, snowline radii are taken as 0.5 $\times$ FWHM of the minor axis of an elliptical Gaussian fit in the image plane. For IRAS16293A, IRAS2A, IRAS4A-NW and IRAS4B, snowline radii are taken as 0.5 $\times$ FWHM of a circular Gaussian fit in the (\textit{u,v})-plane. For IRAS15398, the snowline is estimated from the spatial extent of the spatially resolved central component in the moment zero map.}
\end{deluxetable*}

\end{appendix}

\end{document}